\documentclass[sigconf]{acmart}

\AtBeginDocument{
  }

\usepackage{xcolor}
\definecolor{dullorange}{RGB}{204,119,34}

\usepackage{geometry}
\usepackage{tabularx}
\usepackage{booktabs} 
\usepackage{fancyvrb}
\usepackage{calc}    
\usepackage{graphicx}
\usepackage{algpseudocode}
\usepackage{algorithm}

\usepackage[utf8]{inputenc}
\usepackage{amsmath,amssymb}
\usepackage{makecell}
\usepackage{multirow}
\usepackage{balance}
\usepackage{fvextra} 
\usepackage{longtable}
\usepackage{eso-pic}

\copyrightyear{2025}
\acmYear{2025}
\setcopyright{acmlicensed}
\acmConference[RecSys ’25]{Proceedings of the Nineteenth ACM Conference on Recommender Systems}{September 22--26, 2025}{Prague, Czech Republic}
\acmBooktitle{Proceedings of the Nineteenth ACM Conference on Recommender Systems (RecSys ’25), September 22--26, 2025, Prague, Czech Republic}
\acmDOI{10.1145/3705328.3748064}
\acmISBN{979-8-4007-1364-4/2025/09}

\begin{document}

\title[VL-CLIP: Enhancing Multimodal Recommendations]{VL-CLIP: Enhancing Multimodal Recommendations via \\ Visual Grounding and LLM-Augmented CLIP Embeddings
}

\author{Ramin Giahi}
\authornote{Equal contribution}
\affiliation{
  \institution{Walmart Global Tech}
  \city{Sunnyvale}
  \state{CA}
  \country{USA}
}
\email{ramin.giahi@walmart.com}
\orcid{0000-0002-5261-4069}

\author{Kehui Yao}
\authornotemark[1]
\affiliation{
  \institution{Walmart Global Tech}
  \city{Bellevue}
  \state{WA}
  \country{USA}
}
\email{kehui.yao@walmart.com}
\orcid{0000-0002-7459-795X}

\author{Sriram Kollipara}
\authornotemark[1]
\affiliation{
  \institution{Walmart Global Tech}
  \city{Sunnyvale}
  \state{CA}
  \country{USA}
}
\email{sriram.kollipara@walmart.com}
\orcid{0009-0004-6268-4265}

\author{Kai Zhao}
\authornotemark[1]
\affiliation{
  \institution{Walmart Global Tech}
  \city{Sunnyvale}
  \state{CA}
  \country{USA}
}
\email{kai.zhao@walmart.com}
\orcid{0000-0003-1040-0211}

\author{Vahid Mirjalili}
\authornotemark[1]
\affiliation{
  \institution{Walmart Global Tech}
  \city{Sunnyvale}
  \state{CA}
  \country{USA}
}
\email{vahid.mirjalili@walmart.com}
\orcid{0000-0003-0300-5344}

\author{Jianpeng Xu}
\affiliation{
  \institution{Walmart Global Tech}
  \city{Sunnyvale}
  \state{CA}
  \country{USA}
}
\email{jianpeng.xu@walmart.com}
\orcid{0000-0003-3702-528X}

\author{Topojoy Biswas}
\affiliation{
  \institution{Walmart Global Tech}
  \city{Sunnyvale}
  \state{CA}
  \country{USA}
}
\email{topojoy.biswas@walmart.com}
\orcid{0009-0006-0239-0606}

\author{Evren Korpeoglu}
\affiliation{
  \institution{Walmart Global Tech}
  \city{Sunnyvale}
  \state{CA}
  \country{USA}
}
\email{ekorpeoglu@walmart.com}
\orcid{0009-0003-7754-3652}

\author{Kannan Achan}
\affiliation{
  \institution{Walmart Global Tech}
  \city{Sunnyvale}
  \state{CA}
  \country{USA}
}
\email{kannan.achan@walmart.com}
\orcid{0009-0000-9186-3175}

\renewcommand{\shortauthors}{Ramin Giahi et al.}

\begin{abstract}
Multimodal learning plays a critical role in e-commerce recommendation platforms today, enabling accurate recommendations and product understanding. However, existing vision-language models, such as CLIP, face key challenges in e-commerce recommendation systems: 1) Weak object-level alignment, where global image embeddings fail to capture fine-grained product attributes, leading to suboptimal retrieval performance; 2) Ambiguous textual representations, where product descriptions often lack contextual clarity, affecting cross-modal matching; and 3) Domain mismatch, as generic vision-language models may not generalize well to e-commerce-specific data. To address these limitations, we propose a framework, VL-CLIP, that enhances CLIP embeddings by integrating Visual Grounding for fine-grained visual understanding and an LLM-based agent for generating enriched text embeddings. Visual Grounding refines image representations by localizing key products, while the LLM agent enhances textual features by disambiguating product descriptions. Our approach significantly improves retrieval accuracy, multimodal retrieval effectiveness, and recommendation quality across tens of millions of items on one of the largest e-commerce platforms in the U.S., increasing CTR by 18.6\%, ATC by 15.5\%, and GMV by 4.0\%. Additional experimental results show that our framework outperforms vision-language models, including CLIP, FashionCLIP, and GCL, in both precision and semantic alignment, demonstrating the potential of combining object-aware visual grounding and LLM-enhanced text representation for robust multimodal recommendations.
\end{abstract}

\ccsdesc[500]{Information systems~Recommender systems}
\ccsdesc[500]{Information systems~Users and interactive retrieval}

\keywords{Multimodal Learning, E-Commerce, CLIP, Visual Grounding, Large Language Models, Image-Text Representation, Retrieval, AI for Recommendation}

\maketitle

\section{Introduction}

E-commerce platforms have revolutionized the way consumers interact with products, offering extensive catalogs that cater to diverse preferences. As the number of products continues to grow exponentially, delivering highly relevant personalized recommendations has become an increasingly complex challenge. Consumers often rely on multimodal interactions—searching with a combination of textual queries and images—to find the products they desire. Therefore, improving multimodal representation learning is critical for enhancing search accuracy, recommendation quality, and overall user experience in e-commerce \cite{zhu2024bringing}.

\begin{figure*}[t]
    \centering
    \includegraphics[width=0.90\textwidth]{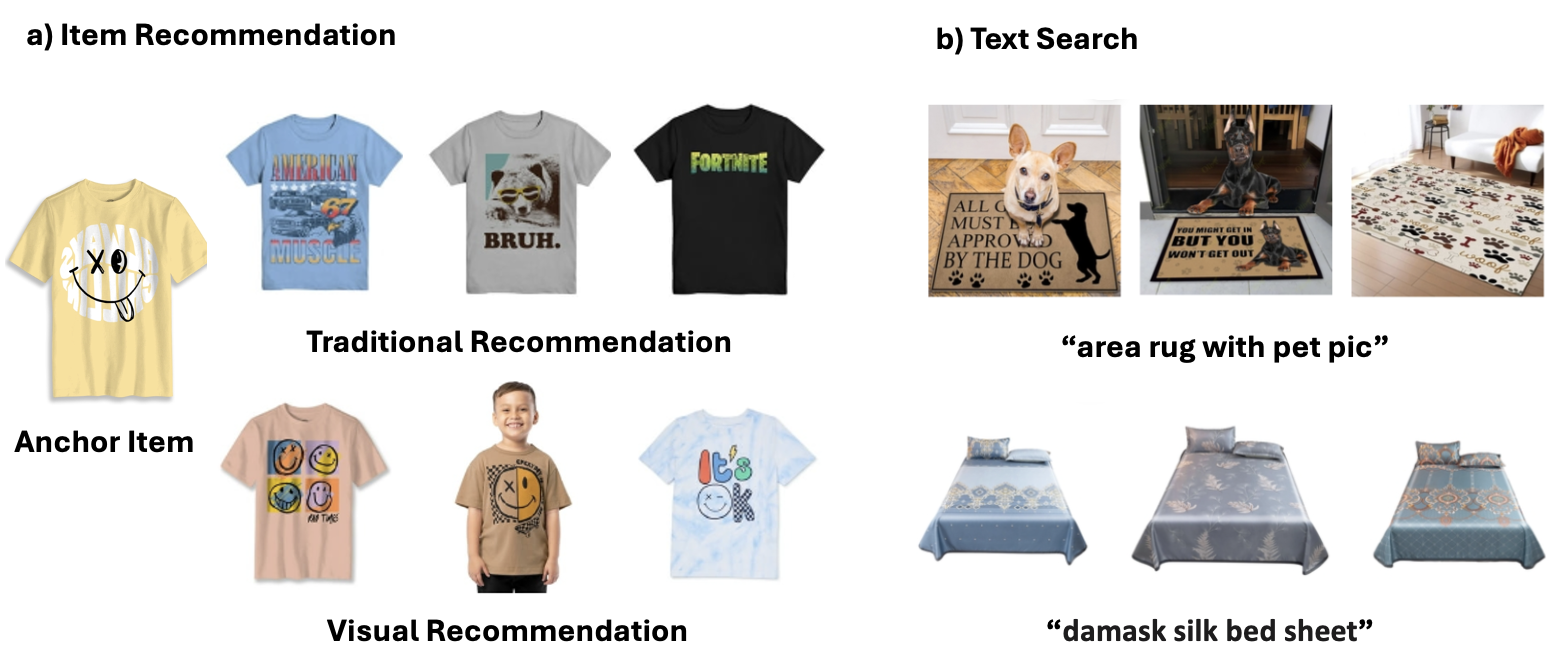}
    \caption{Illustration of: (a) visual recommendation improvement achieved by our proposed model, VL-CLIP and (b) visual search improvement using VL-CLIP.}
    \Description{}
    \label{fig:figure1}
\end{figure*}

Recent advances in vision-language models have significantly improved cross-modal retrieval. CLIP \cite{radford2021learning}, in particular, has demonstrated strong zero-shot capabilities by aligning images and text in a shared embedding space. However, despite its success, CLIP exhibits several  limitations when applied to e-commerce scenarios. First, CLIP processes images globally, meaning that it often fails to capture fine-grained product attributes that are crucial to distinguish visually similar but semantically different items. For example, two handbags might appear nearly identical in a global embedding space, even if one has a unique texture or clasp design that differentiates it. This weak object-level alignment leads to suboptimal retrieval performance, especially in a large e-commerce platform.

Another major challenge is the ambiguity of textual representations. Product descriptions in e-commerce catalogs vary widely in quality and consistency. Some descriptions are too verbose, containing extraneous information, while others are sparse, lacking essential details. CLIP’s text encoder struggles with such inconsistencies, especially with long-text descriptions, leading to poor semantic alignment between textual and visual representations. Without structured and enriched textual inputs, CLIP may misinterpret product intent, reducing the accuracy of multimodal retrieval.

Moreover, existing multimodal models are typically trained on general-purpose datasets, such as LAION-400M \cite{schuhmann2021laion}, which contain a broad spectrum of image-text pairs. While this training paradigm enables broad zero-shot learning, it also introduces a significant domain mismatch when applied to e-commerce. Product images often contain controlled backgrounds, well-lit professional shots, or lifestyle depictions, all of which differ from the diverse, noisy images seen in open-domain datasets. Consequently, pre-trained models fail to generalize effectively to e-commerce-specific data, necessitating domain adaptation strategies \cite{liu2025multimodal}.

To overcome these limitations, we propose a novel framework that enhances CLIP embeddings through two key innovations: (1) the integration of \textbf{Visual Grounding} for fine-grained object localization and (2) the use of a \textbf{Large Language Model (LLM)} to refine textual embeddings. Visual Grounding \cite{liu2025grounding} enables precise localization of key product attributes within an image, ensuring that CLIP’s vision encoder focuses on the most relevant regions. By incorporating Visual Grounding, we improve object-level alignment, leading to more discriminative visual embeddings.

On the textual side, we employ an LLM agent to enrich product descriptions by generating structured, semantically meaningful text representations. Given raw metadata, the LLM refines descriptions, removes noise, and injects domain-specific knowledge, ultimately improving the quality of text embeddings. This augmentation mitigates CLIP’s struggle with ambiguous text and ensures that the image-text alignment is robust, accurate, and context-aware.

Figure~\ref{fig:figure1} illustrates the effectiveness of our approach in both visual and textual recommendation. In Figure~\ref{fig:figure1} (a), the traditional recommendation system suggests products based on broad categorical similarity, often missing fine-grained visual coherence. In contrast, our visual recommendation system, powered by Visual Grounding and enhanced CLIP embeddings, retrieves visually and semantically aligned items, improving recommendation relevance. Similarly, Figure~\ref{fig:figure1} (b) highlights how our model enhances e-commerce search. Traditional keyword-based search may yield inconsistent results when dealing with complex queries such as “\textit{area rug with pet pic}” or “\textit{damask silk bed sheet.}” Our model effectively aligns textual queries with the most relevant visual content, ensuring that search results are not only textually but also visually accurate. These improvements validate our approach’s superiority in capturing fine-grained details and providing semantically meaningful retrievals, ultimately enhancing the user experience.

The contributions of our paper are threefold: First, we introduce a novel multimodal pipeline that integrates Visual Grounding and LLM-enhanced embeddings to improve fine-grained alignment in e-commerce applications; Second, we develop a scalable retrieval and ranking system that efficiently handles large-scale product catalogs; Third, we validate our approach through extensive experiments over tens of millions of items in \textit{Walmart.com}, demonstrating significant improvements in retrieval accuracy, recommendation quality, and overall system performance compared to existing state-of-the-art multimodal models. 

The remainder of this paper is organized as follows. Section 2 discusses related work in multimodal learning, vision-language models, and e-commerce recommendation systems. Section 3 describes our proposed framework, detailing the enhancements to both image and text representations. Section 4 presents experimental results, including comparative evaluations and ablation studies. Section 5 concludes the paper.

\section{Related Work}

Multi-Modality learning has long been an active area of research. The advances in pre-trained vision language models enable applications across diverse domains such as healthcare \cite{mesko2023, huang2023}, finance \cite{dolphin2022multimodal}, social networks \cite{Bell2020GrokNetUC, ofli2020analysis}, search engines \cite{zhai2019learning, dolev2023efficientlargescalevisualrepresentation}, and e-commerce \cite{jin2023learning,ma2024triple}. Transformer-based architectures revolutionized multi-modal learning. By integrating textual and visual input into a unified latent space through self-attention and cross-attention mechanisms, models such as VL-BERT \cite{su2020vlbertpretraininggenericvisuallinguistic}, ViLBERT \cite{lu2019vilbertpretrainingtaskagnosticvisiolinguistic}, and LXMERT \cite{tan2019lxmertlearningcrossmodalityencoder} laid the foundation for robust vision language reasoning. Subsequent models, including VisualBERT \cite{li2019visualbertsimpleperformantbaseline}, UNITER \cite{chen2020uniteruniversalimagetextrepresentation}, and OSCAR \cite{li2020oscarobjectsemanticsalignedpretraining}, further refined these capabilities, achieving state-of-the-art performance across multiple benchmarks and enabling generalized representation learning.

\begin{figure*}[t]
    \centering
    \includegraphics[width=0.85\textwidth]{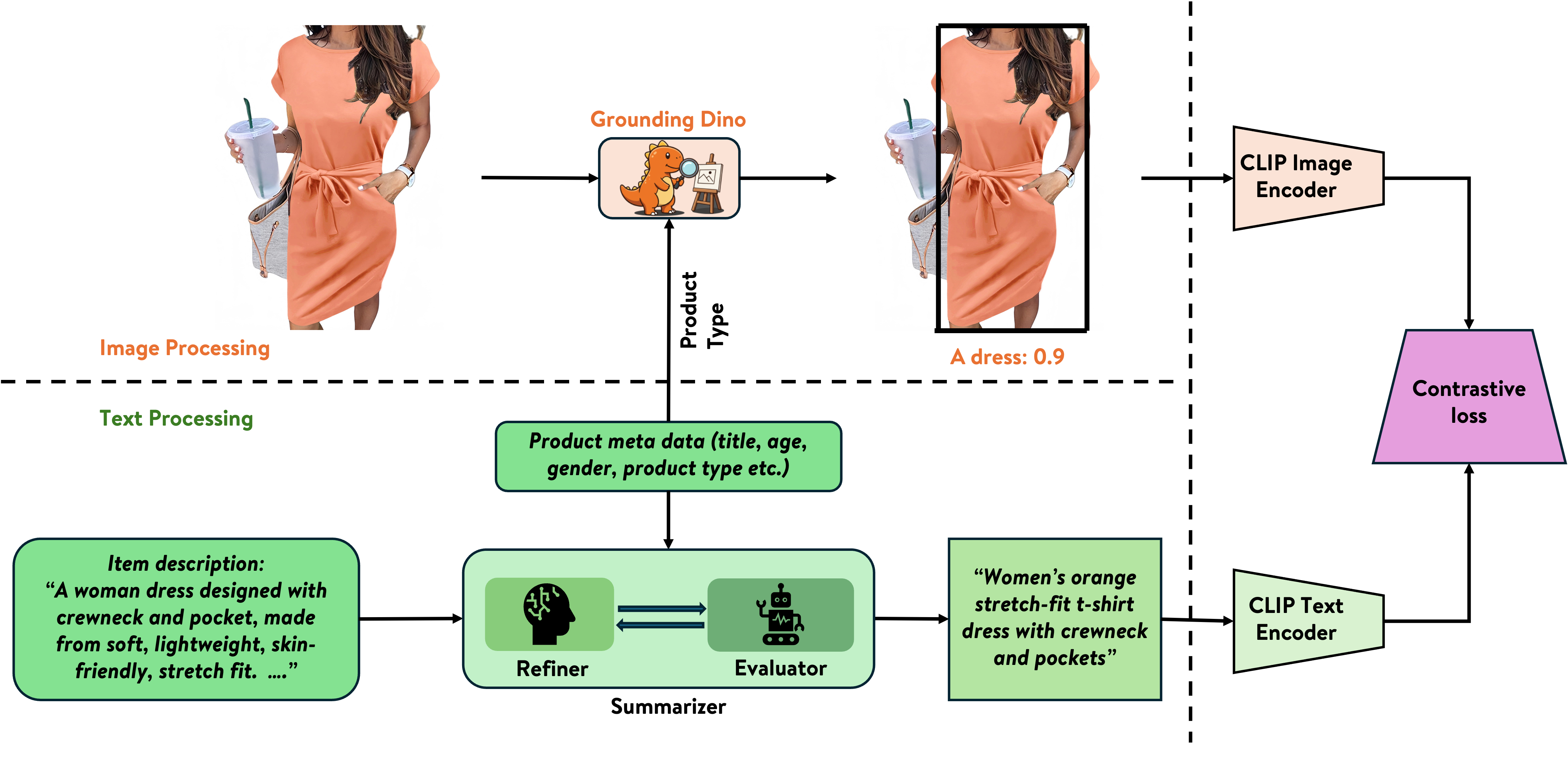}
    \caption{VL-CLIP model architecture}
    \Description{}
    \label{fig:figure2}
\end{figure*}

In parallel to attention-based mechanisms, Radford et al. introduced the CLIP \cite{radford2021learning} model, a dual encoder approach, trained on vast amounts of noisy image-text data. It sparked significant interest by showcasing robust performance across various vision-language tasks. Using contrastive learning mechanism to directly align visual and textual embeddings in shared space, it enabled impressive zero shot retrieval capabilities. Many works have extended CLIP by scaling up data \cite{Cherti_2023}, improving data curation \cite{Cherti_2023, schuhmann2022laion5bopenlargescaledataset}, altering inputs \cite{xu2021videoclipcontrastivepretrainingzeroshot, guzhov2021audioclipextendingclipimage}, refining the loss function or alignment strategy \cite{yao2021filipfinegrainedinteractivelanguageimage, ma2022xclipendtoendmultigrainedcontrastive}, adapting to new tasks \cite{zhong2021regionclipregionbasedlanguageimagepretraining, mokady2021clipcapclipprefiximage}, ranking \cite{zhu2024generalizedcontrastivelearningmultimodal} and domain adaptation \cite{chia2023contrastivelanguagevisionlearning,RSCLIP}. 

Building on the capabilities of CLIP, we fine-tune its dual encoder architecture to adapt to the e-commerce domain, where multi-modal retrieval is critical for matching textual queries to product images. Our approach involves leveraging domain-specific datasets comprising noisy and diverse image-text pairs, a hallmark of e-commerce platforms.
By tailoring CLIP to handle e-commerce-specific challenges, we aim to achieve superior alignment and retrieval performance, ultimately improving customer experience in search and recommendation systems.

\section{Methodology}

In this section, we introduce VL-CLIP, a systematic framework for fine-tuning the CLIP model to achieve robust image-text alignment in e-commerce applications (see Figure \ref{fig:figure2}). The framework integrates advanced vision-language techniques across three stages: 1) image region refinement with Visual Grounding, 2) LLM-driven textual query synthesis, and 3) contrastive training with CLIP optimizations. Below, we provide a comprehensive breakdown of each component, including implementation specifics and design rationale. This robust approach addresses challenges of data noise, domain-specific alignment, and scalability. All the mathematical symbols used in this paper are listed in the table~\ref{tab:symbol} in Appendix~\ref{appendix:nomenclature}.

\subsection{Image Region Refinement with Visual Grounding}

To focus on product-relevant regions, we employed Grounding DINO (GD)—a zero-shot object detection model that aligns visual regions with text prompts \cite{liu2025grounding} for Visual Grounding. For each image, the product type extracted from the product metadata (e.g., ``\textit{dress},'' ``\textit{backpack}'') was used as the text prompt to grounding dino to generate candidate boxes along with confidence scores. The top-scoring box was selected, and its region was cropped and resized. If no box exceeded a confidence threshold the original image was retained to avoid losing critical context. Visual Grounding’s ability to leverage semantic text prompts ensures precise localization of product-centric regions, reducing noise from irrelevant backgrounds (e.g., studio props). To enhance the focus on product-relevant visual elements, we employ the following steps to refine image inputs:

Given an image $I$, Grounding DINO generates a set of $N$ bounding box proposals:

\[
    B = \{ b_1, b_2, \dots, b_N \}
\]

\noindent Each bounding box $b_i \in B$ is associated with a confidence score $s_i$:

\[
    s_i = \frac{\exp(\phi_{\text{image}}(v_i) \cdot \phi_{\text{text}}(P) / \tau_{\text{DINO}})}
    {\sum_{j=1}^{N} \exp(\phi_{\text{image}}(v_j) \cdot \phi_{\text{text}}(P) / \tau_{\text{DINO}})}
\]

\noindent where $\phi_{\text{image}}(v_i)$ and $\phi_{\text{text}}(P)$ represent the Grounding DINO's encoders for the image region $v_i$ and text prompt $P$, $\tau_{\text{DINO}}$ is the temperature parameter, and $s_i$ represents the probability of $b_i$ being the most relevant region. The highest-confidence bounding box $b^*$ is selected using:

\[
    i^* = \arg\max_{i \in \{1, \dots, N\}} s_i
\]
If the confidence score of $b_{i^*}$ is below a pre-defined threshold $\tau_{\text{thresh}}$, the full image is retained:

\[
    I_{\text{crop}} =
    \begin{cases} 
        \text{Crop}(I, b_{i^*}), & \text{if } s_{i^*} \geq \tau_{\text{thresh}} \\
        I, & \text{otherwise}
    \end{cases}
\]

\noindent where $\text{Crop}(I, b_{i^*})$ extracts the product-centered region based on the selected bounding box, and $I_{\text{crop}}$ is the final refined image input. Once the refined image $I_{\text{crop}}$ is obtained, it is passed through the CLIP vision encoder $\phi_{\text{CLIP-image}}$ to obtain its feature embedding:

\[
    v = \frac{\phi_{\text{CLIP-image}}(I_{\text{crop}})}{\|\phi_{\text{CLIP-image}}(I_{\text{crop}})\|}
\]

\noindent where $v$ is the normalized image embedding. By leveraging Visual Grounding for region refinement, we ensure that the extracted embeddings capture fine-grained product attributes, leading to improved alignment in multimodal retrieval.

\subsection{LLM-driven Textual Query Synthesis}

To improve textual representations for multimodal retrieval, we introduce an \textbf{LLM-driven text refinement process}. This process enhances product descriptions by generating structured and semantically rich queries that align better with visual features. The approach consists of three main components: \textit{Summarization, Evaluation, and Refinement}.

Given a raw textual input consisting of both structured and unstructured product information, we first construct an initial concatenated metadata representation as $t_{\text{concat}}$.
\begin{equation}
t_{\text{concat}} = [t_p \parallel t_g \parallel t_{\text{raw}} \parallel t_{\text{in-context}}]\notag
\end{equation}

\noindent where $t_p$ denotes the \textit{product type} (e.g., ``\textit{t-shirt,}'' ``\textit{handbag}''), $t_g$ represents \textit{age and gender attributes} (when applicable), $t_{\text{raw}}$ represents the original \textit{product title and description} and $t_{\text{in-context}}$ contains few-shot examples curated to guide the LLM’s behavior in ambiguous cases. This concatenated information is summarized by an LLM-based summarizer to form the initial query $q^{\text{init}}$. 
\begin{equation}
q^{\text{init}} = \text{Summarizer}(t_{\text{concat}})\notag
\end{equation}
{ Given the recent advances have demonstrated strong few-shot capabilities of LLMs \cite{brown2020gpt3}, we leverage curated set of few-shot examples, specifically designed to address scenarios where LLM exhibits misalignment in $t_{\text{in-context}}$. This allows us to reinforce the desired behavior and improve performance, while maintaining model generality}. 

Next, we iteratively refine this initial query into a structured, concise, and visually relevant query using two specialized LLM-based modules: an Evaluator and a Refiner.

{
Let $\text{Evaluator}(q, t_{\text{concat}})$ be an LLM-based function assessing query $q$'s quality against the concatenated input text $t_{\text{concat}}$ based on these criteria:}
\begin{enumerate}
    \item \textbf{Attribute Consistency}: { Ensures the query reflects attributes present in the input. For example, if $q$ specifies color as red, this criterion evaluates whether the $t_{\text{concat}}$ contains a color attribute and that it is indeed red. }
    \item \textbf{Conciseness}: Limits length of query to \textit{10--20 words} while preserving the meaning.
    \item \textbf{Alignment with Visual Data}: { 
    
    Retains only visually discernible attributes. For example, if the $t_\text{concat}$ mentions a t-shirt is "striped and quick-drying", this criterion ensures we retain only "striped" since it's visually discernible, while excluding "quick-drying" as a non-visual functional attribute.}
\end{enumerate}
The Evaluator outputs either a refinement suggestion or a special token \texttt{<STOP>} when no further improvements are necessary. Let $\text{Refiner}(q, e)$ be an LLM-based function that generates a refined query using the current query $q$ and feedback $e$ from the Evaluator. We denote the Evaluator's output and refined query at iteration $i$ as $e^i$ and $q^i$, respectively.

Starting with $q^{\text{init}}$ as $q^{0}$ here, at each iteration $i$ { ($1 \leq i \leq i_\text{max}$)} , the Evaluator first assesses the query from the previous iteration $q^{i-1}$ and provides feedback $e^i = \text{Evaluator}(q^{i-1}, t_{\text{concat}})$. If the Evaluator indicates that no further improvement is necessary by returning \texttt{<STOP>}, the iterative refinement process terminates, and the query $q^{i-1}$ is accepted as final. Otherwise, the Refiner function uses the Evaluator's feedback to generate an improved query for the next iteration $q^i = \text{Refiner}(q^{i-1}, e^i, t_{\text{concat}})$. {We empirically set $i_{\max} = 5$ as this provides sufficient iterations for convergence while maintaining computational efficiency.}

\begin{figure}[h]
    \centering
    \includegraphics[width=0.50\textwidth]{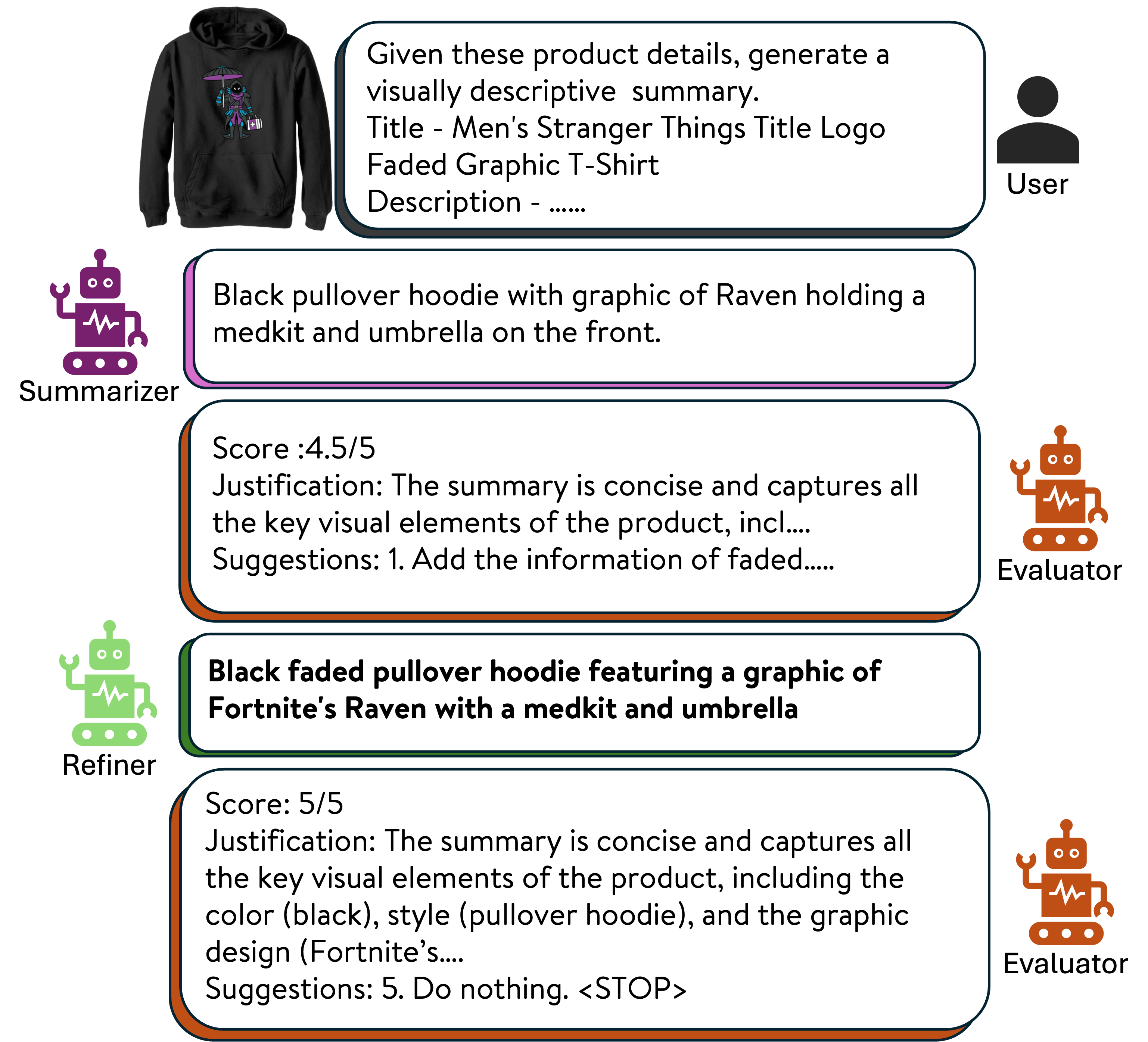}
    \caption{Visualization of product summary generator}
    \Description{}
    \label{fig:LLM_summarizer}
\end{figure}

After the iterative refinement concludes, we obtain the final refined query, denoted as  $q^{\text{final}}$. This query is then embedded into a semantic space suitable for multimodal retrieval by a \textbf{text encoder} $\phi_T$, producing a normalized embedding vector $t$:
\begin{equation}
    t = \frac{\phi_{\text{CLIP-text}}(q^{\text{final}})}{\|\phi_{\text{CLIP-text}}(q^{\text{final}})\|}\notag
\end{equation}

\noindent where $t$ represents the \textbf{normalized textual embedding} used for matching against the \textbf{image embeddings} in the retrieval model. By employing this \textbf{LLM-driven synthesis method}, the textual representations become \textbf{more structured, visually aligned, and domain-adapted}, ultimately enhancing the performance of the multimodal retrieval system. This iterative loop illustrated in Figure \ref{fig:LLM_summarizer}, echoing the self-reflection and self-correction mechanisms, allows the model to autonomously improve its output.

{ The prompts used for the Summarizer, Evaluator, and Refiner are provided in Appendix~\ref{sec:synthesis_prompts}.}

\subsection{Contrastive Fine-tuning of CLIP}

We align image and text embeddings in a shared semantic space to fine-tune CLIP, overcoming general-purpose model limitations. We employ a symmetric contrastive loss function, maximizing similarity between matched image-text pairs while minimizing it for mismatches. This ensures robust alignment across modalities. A fine-tuned $\text{ViT-B}/32$ processes cropped images, while a transformer-based text encoder refines LLM-augmented queries. Both produce 512-dimensional embeddings optimized for e-commerce-specific retrieval tasks. Training involves multiple epochs, leveraging domain-specific augmentations to achieve higher precision in retrieval and classification tasks. This introduces a systematic framework for fine-tuning the CLIP model to achieve robust image-text alignment in e-commerce applications. The symmetric {InfoNCE-style} loss maximizes similarity for matched pairs and minimizes it for negatives:

\begin{equation*}
\mathcal{L}_{\text{CLIP}} = -\frac{1}{2N} \sum_{i=1}^N \left[ \log \frac{e^{v_i \cdot t_i / \tau}}{\sum_{j=1}^N e^{v_i \cdot t_j / \tau}} + \log \frac{e^{t_i \cdot v_i / \tau}}{\sum_{j=1}^N e^{t_i \cdot v_j / \tau}} \right]
\end{equation*}

\noindent where \(\tau \) is the temperature of contrastive loss. We summarize the step-by-step procedure for the VL-CLIP training in Algorithm~\ref{alg:VL-CLIP} in {Appendix~\ref{appendix:training_algo}}.

\subsection{Online Deployment and Scalability}

In this section, we introduce our pipeline and how  we deploy VL-CLIP at scale over tens of millions of shopping items { in Walmart’s e-commerce platform}. The production inference pipeline combines multimodal processing, efficient indexing, and scalable retrieval to provide recommendations for e-commerce applications. In the following, we detail each component, how we scale it, and its role in the system.

\subsubsection{Image and text preprocessing}

We leverage perceptual hashing (pHash)\cite{phash_zauner2010implementation}, a technique that generates compact and robust hash representations of images, which generates fingerprints invariant to resizing and compression. Images were hashed using perceptual hashing techniques to identify and remove duplicates, reducing redundancy in the catalog. After de-duplication,  images are processed by Visual Grounding to crop product-centric regions. This reduces false positives caused by background variations (e.g., the same dress on different mannequins). Visual Grounding dynamically crops product-centric regions using metadata-derived prompts (e.g., ``\textit{handbag}'').

\subsubsection{Hierarchical Navigable Small World (HNSW) index}

Embeddings are indexed using HNSW\cite{hnsw_malkov2018efficient}, a graph-based {Approximate Nearest Neighbors} (ANN) algorithm optimized for high recall and low latency. The hierarchical graph structure allows logarithmic-time search complexity. Metadata (e.g., product type) is fused with cropped images to create a unified dataset. This ensures retrieval accounts for both visual and contextual signals. Instead of computing image embeddings for all images in the catalog, we maintain an image embedding database. Generating embeddings for large-scale e-commerce data items at million level is computationally intensive. To handle this, we distribute the workload across multiple machines, each equipped with a T4 GPU.

\subsubsection{Retrieval and pairwise ranking}
For a query embedding $e$, the HNSW index retrieves top-$k$ candidates using cosine similarity. The ANN index was queried to retrieve visually similar items. Efficient index construction and retrieval are crucial for real-time performance. We optimized the process by grouping items based on product type and constructing separate indices for each group.

\subsubsection{Scalability}

The architecture developed in this work is now fully deployed on Walmart’s e-commerce platform, supporting real-time recommendations and multimodal retrieval at scale. The pipeline integrates data preprocessing, embedding generation, and retrieval in a seamless workflow. { These optimizations reduce search space and memory usage while preserving quality. pHash improves MRR by +7.2\%; product type-based HNSW indexing improves Precision@1 by +9\% and reduces latency by +81\% compared to IVF indexing.}  
Algorithm~\ref{alg:VL-CLIPd} in {Appendix~\ref{appendix:deployment_algo}} shows the inference procedure.

\section{Experiments}

\subsection{Data Preparation}

Millions of product images and metadata (e.g., descriptions, titles, attributes) are sourced from an extensive e-commerce catalog. This diverse dataset includes apparel and home goods, ensuring comprehensive representation of categories. Each sample includes product images, which may be high-quality but could contain distracting elements in the background, such as real-life settings or lifestyle scenes, as well as textual metadata, which consists of structured attributes (product type, gender, age group) and unstructured data (titles, descriptions).

We leverage following pre-processing steps to clean the input data: 1) Image Normalization: Resized the images and normalized using CLIP’s preprocessing pipeline $I_{norm} = \dfrac{I_{resized}-\mu}{\sigma}$, where $\mu$ and $\sigma$ are channel-wise mean and standard deviation values. 2) Text Sanitization: Removed HTML tags, special characters, and redundant keywords from metadata. Descriptive keywords are retained, while noise (e.g., ``\textit{free shipping}'') is excluded, yielding semantically rich inputs. 3) Category Balancing: Stratified sampling ensured proportional representation of product types to mitigate bias that can skew model predictions toward overrepresented categories.

We fine-tune the VL-CLIP model using 7 million products from the fashion and home categories of \textit{Walmart.com} using model architecture described in Figure \ref{fig:figure2}. We evaluated our model on a dataset containing fashion and home items. To ensure variety, we sampled items equally across different product types—such as T-shirts, dresses, and coffee tables—resulting in 10 product types for fashion and 7 for home, for a total of 17 product types. In total, we obtained 10,000 samples for fashion category and 10,000 samples for home category for evaluation.

\subsection{Evaluation Metrics}
The performance of VL-CLIP is compared with existing methods including CLIP~\cite{radford2021learning}, GCL~\cite{zhu2024generalizedcontrastivelearningmultimodal}, and FashionCLIP~\cite{chia2023contrastivelanguagevisionlearning} on multimodal retrieval task on Walmart data. CLIP is a foundational model that learns joint representations from large-scale image–text pairs through contrastive learning~\cite{radford2021learning}. GCL is a generalization of contrastive learning framework that incorporates ranking information alongside multiple input fields containing image-text pairs and queries ~\cite{zhu2024generalizedcontrastivelearningmultimodal}. FashionCLIP is a specialized adaptation of the CLIP paradigm designed for the fashion domain, leveraging fine-grained annotations and domain-specific features~\cite{chia2023contrastivelanguagevisionlearning}.

We measure retrieval performance using two standard metrics:

\begin{itemize}
    \item \textbf{HITS@k:} This metric reports the fraction of queries for which the correct item is among the top $k$ results in the ranked list. Formally, for $N$ queries, each query $i$ has a ground-truth correct item $c_i$. After ranking all items according to a similarity score, let $\mathrm{rank}(c_i)$ be the position of $c_i$. Then $\mathrm{HITS@k} = \frac{1}{N} \sum_{i=1}^{N} \mathbf{1}\bigl(\mathrm{rank}(c_i) \leq k\bigr)$,
    where $\mathbf{1}(\cdot)$ is an indicator function that returns 1 if $\mathrm{rank}(c_i) \leq k$ and 0 otherwise. In our evaluation, we use HITS@5.
    
    \item \textbf{Mean Reciprocal Rank (MRR):} For a query $i$, if the correct item $c_i$ is ranked at $\mathrm{rank}(c_i)$, its reciprocal rank is $\mathrm{RR}_i = \frac{1}{\mathrm{rank}(c_i)}$.    
    The MRR is the average of these reciprocal ranks across all $N$ queries, given by $\mathrm{MRR} = \frac{1}{N} \sum_{i=1}^{N} \mathrm{RR}_i$. This metric particularly favors correct items that rank higher in the list.
\end{itemize}

\subsection{Retrieval Results}

Table \ref{tab:multi-modal-retrieval-performance} illustrates how CLIP, GCL, FashionCLIP, and our proposed VL-CLIP perform on the Fashion and Home datasets, using the HITS@5 and MRR metrics. CLIP, the baseline pre-trained vision-language model, shows modest retrieval capability (HITS@5 of 0.3080 on Fashion and 0.2355 on Home), likely because its global embeddings struggle to capture fine-grained product attributes. The multi-modal retrieval task involves identifying the most relevant image from a given set based on a textual description. For example, in a product retrieval scenario, the goal is to match a product description with its corresponding image in a catalog. 

GCL improves upon CLIP by integrating fine-grained relevance scores into the contrastive learning process, allowing it to explicitly learn nuanced ranking signals rather than binary relevance alone, thus achieving higher metrics (HITS@5 of 0.3992 on Fashion and 0.3104 on Home). However, its reliance on ranking information alone does not fully address domain-specific nuances in product images and textual descriptions.

FashionCLIP {further improves the} performance (HITS@5 of 0.4428 on Fashion and 0.4227 on Home) by applying domain adaptation strategies optimized for fashion. This adaptation allows the model to better encode style and design elements that are particularly relevant for apparel, yet it also provides a notable boost on the Home dataset, indicating that fine-tuning vision-language representations with domain-aware features can generalize beyond the original domain.

VL-CLIP delivers the highest retrieval accuracy and ranking quality across both datasets, as demonstrated by its leading HITS@5 and MRR scores (0.6758 and 0.5252 on Fashion, and 0.6692 and 0.5100 on Home). By integrating local object-level grounding for visual representations and large language model–enriched text embeddings, VL-CLIP captures key product details and resolves ambiguous textual descriptions more effectively than competing methods. The result is a more precise alignment between images and text, which proves especially valuable in e-commerce scenarios where seemingly subtle product attributes and nuanced language can critically impact retrieval success.

\begin{table}[h]
    \centering
    \caption{Multi-modal retrieval performance of different models on Fashion and Home datasets}
    \begin{tabular}{l cc cc}
        \toprule
        \multirow{2}{*}{\textbf{Model}} & \multicolumn{2}{c}{\textbf{Fashion}} & \multicolumn{2}{c}{\textbf{Home}} \\
        
        \cmidrule(lr){2-3} \cmidrule(lr){4-5}
        & \textbf{HITS@5} & \textbf{MRR} & \textbf{HITS@5} & \textbf{MRR} \\
         \midrule
        CLIP  & 0.3080 & 0.2387 & 0.2355 & 0.1747\\
        GCL   & 0.3992 & 0.2952 & 0.3104 & 0.2312\\
        FashionCLIP & 0.4428 & 0.3555 & 0.4227 & 0.3219\\
        
        VL-CLIP & \textbf{0.6758} & \textbf{0.5252} & \textbf{0.6692} & \textbf{0.5100}\\
        \bottomrule
    \end{tabular}
    
    \label{tab:multi-modal-retrieval-performance}
\end{table}

\subsection{Ablation Study}

{To gain deeper insight into the role of each component within the VL-CLIP framework, we perform an ablation study by eliminating essential modules, Visual Grounding and LLM-based query refinement, and assessing how their removal affects retrieval performance.}

{ The results of this ablation analysis are summarized in Table~\ref{tab:ablation_study}}. The full VL-CLIP model achieves the highest performance with a HITS@5 of 0.6758 and an MRR of 0.5252. Removing Visual Grounding results in an average performance drop of 15.34\% in HITS@5 and 11.23\% in MRR across the Fashion and Home categories, demonstrating the importance of background removal and focusing on the main item in enhancing visual matching. Additionally, removing the LLM-based query refinement step further reduces performance by 7.40\% in HITS@5 and 5.32\% in MRR when compared to the model already lacking Visual Grounding, indicating that refining text queries improves retrieval accuracy by providing clearer and more precise textual descriptions. This ablation study highlights that both Visual Grounding and LLM-based query enhancement play crucial roles in improving retrieval effectiveness.

\begin{table}[h]
    \centering
    \caption{Ablation study on the contribution of each component for Fashion and Home datasets}
    \small
    \renewcommand{\arraystretch}{1.1} 
    \begin{tabular}{lcc cc}
        \toprule
        \multirow{2}{*}{\textbf{Model Variant}} & \multicolumn{2}{c}{\textbf{Fashion}} & \multicolumn{2}{c}{\textbf{Home}} \\
        \cmidrule(lr){2-3} \cmidrule(lr){4-5}
        & \textbf{HITS@5} & \textbf{MRR} & \textbf{HITS@5} & \textbf{MRR} \\
        \midrule
 VL-CLIP w/o GD, LLM & 0.4484 & 0.3570 & 0.4418 & 0.3471\\
VL-CLIP w/o GD & 0.5308 & 0.4176 & 0.5075 & 0.3929 \\
 VL-CLIP & \textbf{0.6758} & \textbf{0.5252} & \textbf{0.6692} & \textbf{0.5100} \\
        \bottomrule
    \end{tabular}
    
    \label{tab:ablation_study}
\end{table}

\subsection{Zero-shot Classification}

In addition to the information retrieval and the ablation test, we also performed a zero-shot classification task. We performed two fashion item attribute classification tasks: neckline classification and pattern classification. For neckline classification, we manually selected 1,000 fashion items, each belonging to one of the following categories: v-neck, crew neck, scoop neck, Henley, mock neck, and boat neck. We use a zero-shot classification approach, where we generate a descriptive text for each class (e.g., ``\textit{a T-shirt with a scoop neckline}'') and pass it through a text encoder. The classification is then performed by comparing the image embedding with these text embeddings to find the closest match, which determines the predicted class. Similarly, for pattern classification, we apply the same zero-shot approach using the following categories: ``\textit{solid,}'' ``\textit{cartoon character,}'' ``\textit{heart symbol,}'' and ``\textit{floral print}''.

Table \ref{tab:zero_shot_classification_accuracy} presents the model accuracy for both classification tasks. VL-CLIP consistently outperforms other models, making it the most reliable choice for fashion attribute zero-shot classification. Its superior performance is due to Visual Grounding's ability to remove noise and the LLM-refined queries, which enhance the quality of text-image alignment.

\begin{table}[h]
    \centering
    \caption{Zero-shot performance on pattern and neckline classification tasks}
    \renewcommand{\arraystretch}{1.1} 
    \begin{tabular}{lccc}
        \toprule
        \textbf{Model} & \makecell{\textbf{Neckline} \\ \textbf{classification} \\ \textbf{accuracy}} & \makecell{\textbf{Pattern} \\ \textbf{classification} \\ \textbf{accuracy}}\\
        \midrule
        CLIP  & 0.580 & 0.144  \\
        GCL   & 0.674 & 0.785  \\
        FashionCLIP & 0.881 & 0.934  \\
        VL-CLIP & \textbf{0.937} & \textbf{0.959} \\
        \bottomrule
    \end{tabular}
    
    \label{tab:zero_shot_classification_accuracy}
\end{table}
\subsection{VLM-Agent Evaluation}

Since the alignment of text and image information is very subjective, we employ a VLM agent for evaluation. Our evaluation consists of two retrieval tasks: query-based retrieval and similar item recommendation. The query-based retrieval specifically targets fine-detailed product attributes to ensure accurate retrieval of nuanced product characteristics. For example, ``\textit{Teal floral print blouse}'' is looking for items that match color and pattern characteristics; ``\textit{Beige V-neck short-sleeve T-shirt}''  is looking for color, neckline and sleeve characteristics. For query-based evaluation, the retrieved images corresponding to each query are individually paired with the query and passed to a VLM. The VLM model is asked to assess whether the provided image accurately matches the given query, producing a binary output of 0 (no match) or 1 (match). Similarly, for similar item evaluation, the retrieved images are individually paired with the anchor image, and the VLM is asked to assess whether the two images match in terms of their visual characteristics. We assess the effectiveness of our approach using an VLM-as-judge evaluation framework. More details on this process for automated query generation and VLM-evaluation are provided in Appendix~\ref{appendix:eval_process}.

Table \ref{tab:llm_evaluation} presents the query-based retrieval performance and similar item recommendation performance for \textit{Walmart.com E-Commerce Dataset}. Performance is reported using Precision@1, 3, 5. 
The results show that our VL-CLIP model perform better than the benchmarks including CLIP, FashionClip, and GCL.  Note here the highest values for VL-CLIP appear at Precision@1, gradually decreasing for Precision@3 and Precision@5. This pattern indicates that its top-ranked item is almost always relevant, while subsequent positions, though still relevant, can exhibit slightly lower relevance. In contrast, models like CLIP sometimes show a reversed pattern—with lower Precision@1 than Precision@5—suggesting that their top recommendation is not always the best match, even though they do include relevant items in lower-ranked positions. Examples of both query based and Similar Item (SI) recommendation  tasks are provided in Appendix~\ref{appendix:retrieval_examples}.

The improvements can be attributed to the complementary roles of Visual Grounding and LLM in refining the retrieval process. 
Visual Grounding helps the model focus on the main item within the image, filtering out background distractions, and ensuring that fine-detailed product attributes are emphasized. Meanwhile, LLM enhances the quality of search queries by making them more structured and aligned with real-world user intent. Together, these enhancements enable for more accurate retrieval of products that match specific attribute-based queries.

\begin{table}[h]
    \centering
    \caption{{VLM}-evaluation results of query-based retrieval and similar item recommendation}
    \renewcommand{\arraystretch}{1.1}
    \begin{tabular}{lccc}
        \toprule
        \multicolumn{4}{c}{\textbf{Query-based retrieval}} \\
        \midrule
        \textbf{Model} & \textbf{Precision@1} & \textbf{Precision@3} & \textbf{Precision@5} \\
        \midrule
        CLIP  & 0.3800  & 0.4500  & 0.4620  \\
        FashionCLIP & 0.5900  & 0.6833  & 0.7100  \\
        GCL   & 0.4500  & 0.4800  & 0.4880  \\
        VL-CLIP & \textbf{0.8586}  & \textbf{0.7710}  & \textbf{0.7515}  \\
        \midrule
        \multicolumn{4}{c}{\textbf{Similar item recommendation}} \\
        \midrule
        \textbf{Model} & \textbf{Precision@1} & \textbf{Precision@3} & \textbf{Precision@5} \\
        \midrule
        CLIP  & 0.9719  & 0.9046  & 0.8680 \\
        FashionCLIP & 0.9813  & 0.9582  & 0.9439 \\
        GCL   & 0.9813  & 0.9576  & 0.9349 \\
        VL-CLIP & \textbf{0.9925}  & \textbf{0.9838}  & \textbf{0.9783} \\
        \bottomrule
    \end{tabular}
    
    \label{tab:llm_evaluation}
\end{table}

\subsection{Computation Efficiency}

\begin{figure}
    \centering
    \includegraphics[width=0.40\textwidth]{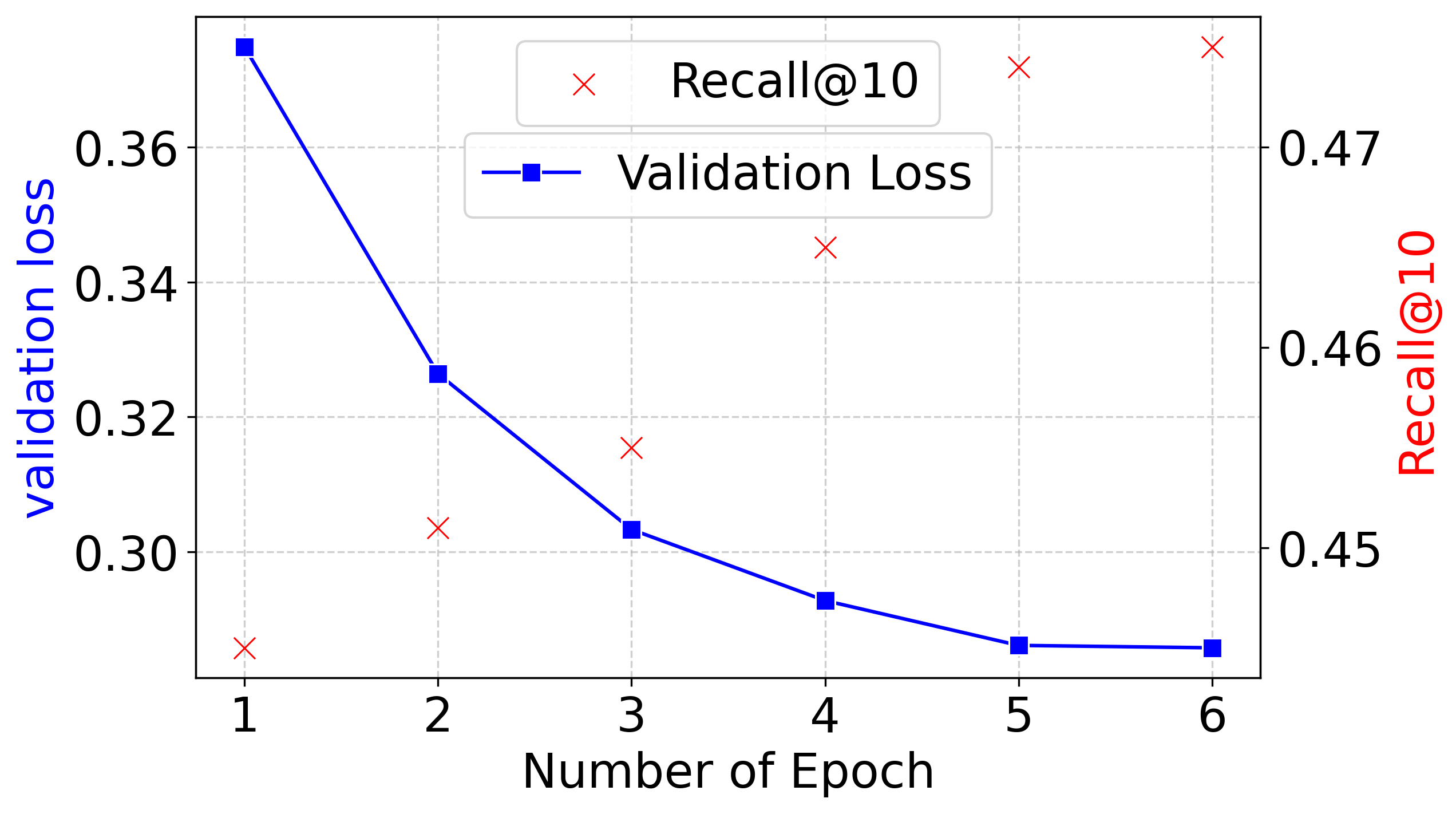}
    \caption{The validation loss and Recall@10 over epochs}
    \Description{}
    \label{fig:training}
\end{figure}

VL-CLIP is fine-tuned on millions of products from the fashion and home categories of \textit{Walmart.com}. Stratified sampling method is applied to ensure proportional representation of diverse set of product types (more than 500 product types). VL-CLIP achieved robust performance on the e-commerce retrieval task over 6 epochs before early stopping (see Figure \ref{fig:training}). The model demonstrated strong alignment between visual and textual embeddings, evidenced by a steady reduction in the contrastive loss for the validation set from 0.38 to 0.28. Retrieval performance, measured by $Recall@10$ indicating that the model effectively identified relevant items in the top-10 results for 47\% of queries. Prolonged training beyond this point led to a marginal decline in Recall@10, suggesting overfitting to noisy pairs or saturation in learning capacity. This underscored the importance of early stopping, with epoch 6 representing the optimal checkpoint for deployment. These results validate the effectiveness of our pipeline—combining Visual Grounding, LLM, and contrastive loss—for scalable e-commerce recommendation systems.

\begin{figure*}[h]
    \centering
    \includegraphics[width=0.90\textwidth]{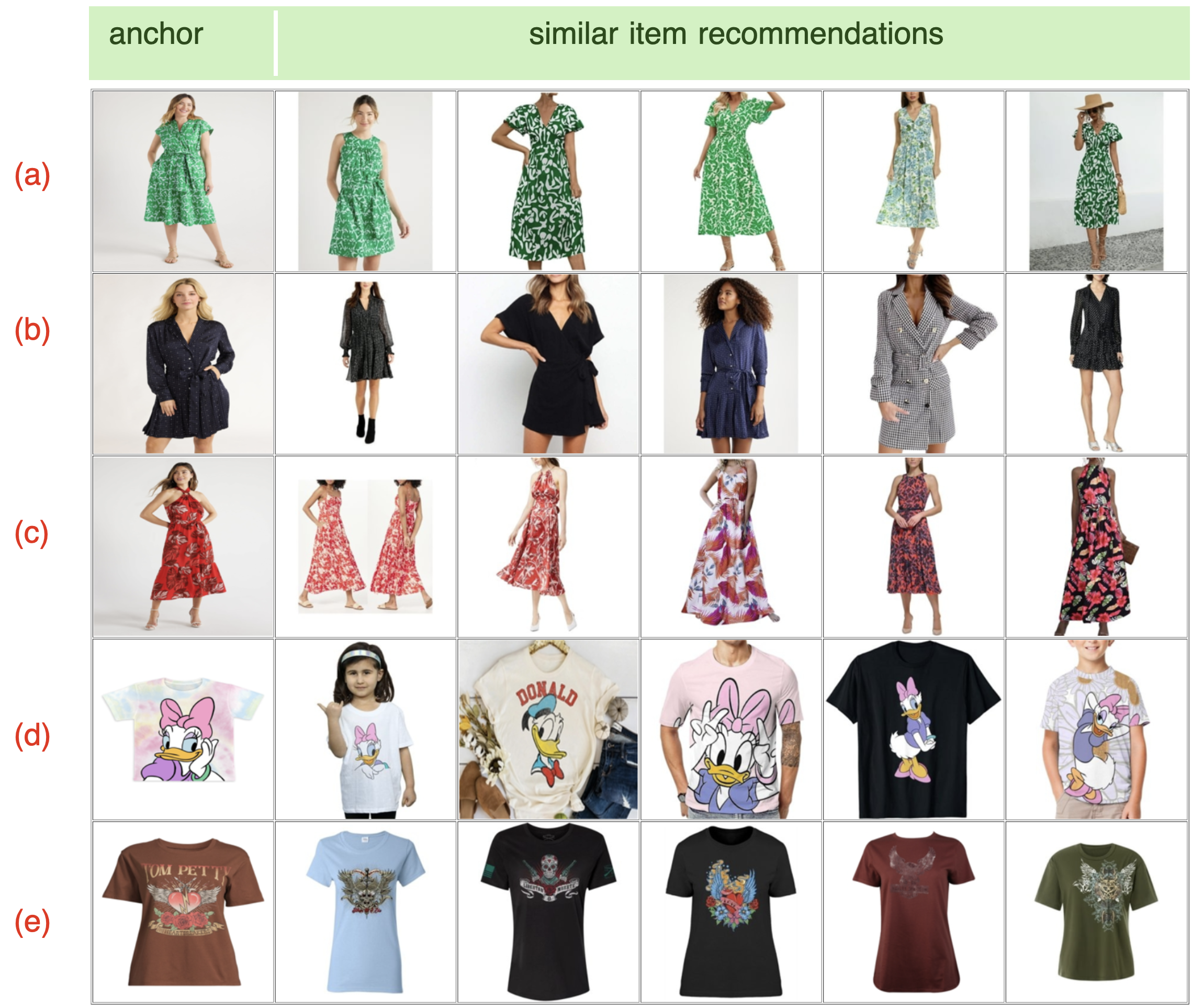}
    \caption{Examples of similar item recommendations for fashion products based on VL-CLIP.}
 \Description{}
    \label{fig:si_fashion}
\end{figure*}

{\subsection{Cross-Domain Generalization}
To assess the generalizability of VL-CLIP, we conduct zero-shot evaluations on a public Google Shopping dataset\footnote{\url{https://github.com/marqo-ai/GCL}}. This data set spans various e-Commerce categories and provides a robust benchmark for testing the model’s ability to transfer knowledge to unseen domains without additional fine-tuning.
It is specifically designed for training and benchmarking multi-modal retrieval models in fine-grained ranking tasks. As shown in Table~\ref{tab:multi-modal-retrieval-performance-google} and Table~\ref{tab:llm_evaluation_google}, VL-CLIP consistently outperforms other models when applied to this new dataset.

We further evaluate zero-shot performance on the \textit{Art} and \textit{Toys} categories from \textit{Walmart.com}, where VL-CLIP again achieves superior results compared to other models. These findings highlight the model’s strong transferability to novel product domains (see Appendix~\ref{appendix:generalizability}).

\begin{table}[h]
    \centering
    \caption{Multi-modal retrieval performance of different models on Google Shopping dataset}
    \begin{tabular}{l cc}
        \toprule
        \textbf{Model} & \textbf{HITS@5} & \textbf{MRR} \\
        \midrule
        CLIP         & 0.2419 & 0.1714 \\
        FashionCLIP  & 0.4495 & 0.3075 \\
        GCL          & 0.6270 & 0.4283 \\
        VL-CLIP      & \textbf{0.6644} & \textbf{0.4936} \\
        \bottomrule
    \end{tabular}
    \label{tab:multi-modal-retrieval-performance-google}
\end{table}

\begin{table}[h]
    \centering
    \caption{{VLM}-evaluation results of query-based retrieval and similar item recommendation on Google Shopping dataset}
    \renewcommand{\arraystretch}{1.1}
    \begin{tabular}{lccc}
        \toprule
        \multicolumn{4}{c}{\textbf{Query-based retrieval}} \\
        \midrule
        \textbf{Model} & \textbf{Precision@1} & \textbf{Precision@3} & \textbf{Precision@5} \\
        \midrule
        CLIP         & 0.3935 & 0.4258 & 0.4167 \\
        FashionCLIP  & 0.7032 & 0.7182 & 0.7238 \\
        GCL          & 0.4193 & 0.4107 & 0.4129 \\
        VL-CLIP      & \textbf{0.8452} & \textbf{0.8215} & \textbf{0.7896} \\
        \midrule
        \multicolumn{4}{c}{\textbf{Similar item recommendation}} \\
        \midrule
        \textbf{Model} & \textbf{Precision@1} & \textbf{Precision@3} & \textbf{Precision@5} \\
        \midrule
        CLIP         & 0.6161 & 0.5684 & 0.5423 \\
        FashionCLIP  & 0.8298 & 0.7980 & 0.7796 \\
        GCL          & 0.7759 & 0.7434 & 0.7141 \\
        VL-CLIP      & \textbf{0.9294} & \textbf{0.9073} & \textbf{0.8950} \\
        \bottomrule
    \end{tabular}
    \label{tab:llm_evaluation_google}
\end{table}

}

\subsection{Online A/B Test}

To validate the effectiveness of VL-CLIP model, we conduct a large-scale A/B test on one of the top two e-commerce platforms in US. The experiment compared the performance of our VL-CLIP with the deployed baseline model. The test lasted four weeks and included millions of user interactions in various product categories. The following key metrics are evaluated in the AB test: Click-Through Rate (CTR), the proportion of users who clicked on recommended products after viewing them; Add-to-Cart Rate (ATC), the percentage of users who added a recommended product to their cart; Gross Merchandise Value (GMV), the total sales revenue generated by the recommendations.

Table \ref{tab:abtest} highlights the relative improvements of our system compared to the baseline model. Online A/B tests validated the effectiveness of VL-CLIP, revealing an 18.6\% increase in CTR, a 15.5\% increase in ATC rate, and a 4\% boost in GMV, underscoring the VL-CLIP’s practical efficacy.  These results highlight the performance of VL-CLIP in understanding user intent and aligning recommendations with user preferences.

\begin{table}[t]
\caption{Online A/B test results}
\begin{tabular}{ccl}
\toprule
\multicolumn{1}{c}{\bf Performance metric}  &\multicolumn{1}{c}{\bf Relative improvement} \\
\midrule
CTR (\%)         &18.6\% \\
ATC(\%)            &15.5\% \\
GMV(\%)           &4.0\%\\
\bottomrule
\end{tabular}
\label{tab:abtest}
\end{table}

We show some case studies in Figure \ref{fig:si_fashion}. The first column is the anchor item, the rest are top five recommended items based on VL-CLIP. In Figure \ref{fig:si_fashion}(a), the anchor item is a green floral midi dress. VL-CLIP retrieves similar style dresses, capturing variations in pattern and length while maintaining the overall aesthetic. Figure \ref{fig:si_fashion}(b) item is a black wrap-style dress with long sleeves. VL-CLIP recommends items with similar sleeve lengths and structured silhouettes, focusing on both color and style.
Figure \ref{fig:si_fashion}(c), (d), and (e) demonstrate the strong fashion understanding capability of VL-CLIP. For further case studies, {please refer to Figure \ref{fig:query_fashion}-\ref{fig:si_home} in Appendix~\ref{appendix:retrieval_examples}}.

\section{Conclusion and Future Work}

In this work, we addressed critical challenges in multimodal representation learning for e-commerce by introducing VL-CLIP, a novel framework that integrates Visual Grounding for visual representation enhancement and LLM-augmented text embeddings for semantic enrichment. Through extensive experiments on large scale e-commerce datasets, VL-CLIP demonstrated superior performance over state-of-the-art baselines. Specifically, HITS@5 improved by 184.16\% on Home dataset and by 119.42\% on the Fashion dataset. 
Furthermore, LLM evaluation results indicate a 62.66\% increase for query-based retrieval and a 12.71\% improvement in similar item recommendations.
Online A/B tests further validated the effectiveness of VL-CLIP, revealing an 18.6\% increase in CTR, a 15.5\% increase in ATC rate, and a 4\% boost in GMV, underscoring the VL-CLIP’s practical efficacy. 
Deploying VL-CLIP on \textit{Walmart.com} highlighted its scalability and real-world impact. The framework’s hierarchical indexing and distributed computation pipeline efficiently processed millions of catalog items.

\newpage

\bibliographystyle{ACM-Reference-Format}
\bibliography{0sample-base}

\appendix

\twocolumn[{
\begin{center}
\vspace{1.5em}
\Huge{\textbf{\Huge VL-CLIP: Enhancing Multimodal Recommendations via\\
Visual Grounding and LLM-Augmented CLIP Embeddings:\\ Appendix}}

\vspace{1.5em}
\normalsize

\begin{minipage}{0.3\textwidth}\centering
Ramin Giahi$^{*}$\\
Walmart Global Tech\\
Sunnyvale, CA, USA\\
\texttt{ramin.giahi@walmart.com}
\end{minipage}
\hfill
\begin{minipage}{0.3\textwidth}\centering
Kehui Yao$^{*}$\\
Walmart Global Tech\\
Bellevue, WA, USA\\
\texttt{kehui.yao@walmart.com}
\end{minipage}
\hfill
\begin{minipage}{0.3\textwidth}\centering
Sriram Kollipara$^{*}$\\
Walmart Global Tech\\
Sunnyvale, CA, USA\\
\texttt{sriram.kollipara@walmart.com}
\end{minipage}

\vspace{1.5em}

\begin{minipage}{0.3\textwidth}\centering
Kai Zhao$^{*}$\\
Walmart Global Tech\\
Sunnyvale, CA, USA\\
\texttt{kai.zhao@walmart.com}
\end{minipage}
\hfill
\begin{minipage}{0.3\textwidth}\centering
Vahid Mirjalili$^{*}$\\
Walmart Global Tech\\
Sunnyvale, CA, USA\\
\texttt{vahid.mirjalili@walmart.com}
\end{minipage}
\hfill
\begin{minipage}{0.3\textwidth}\centering
Jianpeng Xu\\
Walmart Global Tech\\
Sunnyvale, CA, USA\\
\texttt{jianpeng.xu@walmart.com}
\end{minipage}

\vspace{1.5em}

\begin{minipage}{0.3\textwidth}\centering
Topojoy Biswas\\
Walmart Global Tech\\
Sunnyvale, CA, USA\\
\texttt{topojoy.biswas@walmart.com}
\end{minipage}
\hfill
\begin{minipage}{0.3\textwidth}\centering
Evren Korpeoglu\\
Walmart Global Tech\\
Sunnyvale, CA, USA\\
\texttt{ekorpeoglu@walmart.com}
\end{minipage}
\hfill
\begin{minipage}{0.3\textwidth}\centering
Kannan Achan\\
Walmart Global Tech\\
Sunnyvale, CA, USA\\
\texttt{kannan.achan@walmart.com}
\end{minipage}
\end{center}
}

\vspace{4em}
]

\section{Nomenclature}
\label{appendix:nomenclature}
This section presents a table of nomenclature with definitions and explanations of the mathematical symbols used throughout the paper.

\begin{table}[h]
    \caption{Notation and symbols used in this paper}
    \label{tab:symbol}
    \begin{tabular}{ll}
        \toprule
        \multicolumn{1}{l}{\bf Symbol}  &\multicolumn{1}{l}{\bf Definition} \\
        \midrule
        $I$ & Product images  \\
        $t_{\text{raw}}$ & Raw textual metadata (e.g., titles, descriptions) \\
        $t_p$ & Product type (structured attribute, e.g., ``dress'', ``rug'') \\
        $I_{\text{norm}}$ & Normalized (resized) input image \\
        $I_{\text{crop}}$ & Cropped image from GD's top-scoring bounding box \\
        $B$ & Set of bounding boxes from Visual Grounding \\
        $b_{i}$ & $i$-th bounding box \\
        $s_{i}$ & Confidence score for bounding box $b_{i}$ \\
        $\tau$ & temperature of the contrastive loss \\
        $\tau_{\mathrm{DINO}}$ & Temperature parameter for Visual Grounding \\
        $\tau_{\mathrm{thresh}}$ & Confidence threshold for accepting a bounding box \\
        $q^i$ & Refined query at iteration $i$ \\
        $e^i$ & Evaluator's feedback at iteration $i$ \\
        $\phi_{\text{CLIP-image}}$ & CLIP image encoder \\
        $\phi_{\text{CLIP-text}}$  & CLIP text encoder \\
        $v$\ & Image embedding from CLIP image encoder \\
        $t$ & Text embedding from CLIP text encoder \\
        $\tau$ & Temperature parameter in CLIP's contrastive loss \\
        $\mathcal{L}_{\text{CLIP}}$ & Symmetric contrastive loss function for CLIP \\
        $N_{\text{epochs}}$ & Maximum number of training epochs \\
        $H$ & HNSW index \\
        \bottomrule
    \end{tabular}
\end{table}

\section{Visualization for Query-based retrieval and Similar item recommendation task}
\label{appendix:retrieval_examples}

\begin{figure*}[h]
    \centering
    \includegraphics[width=0.98\textwidth]{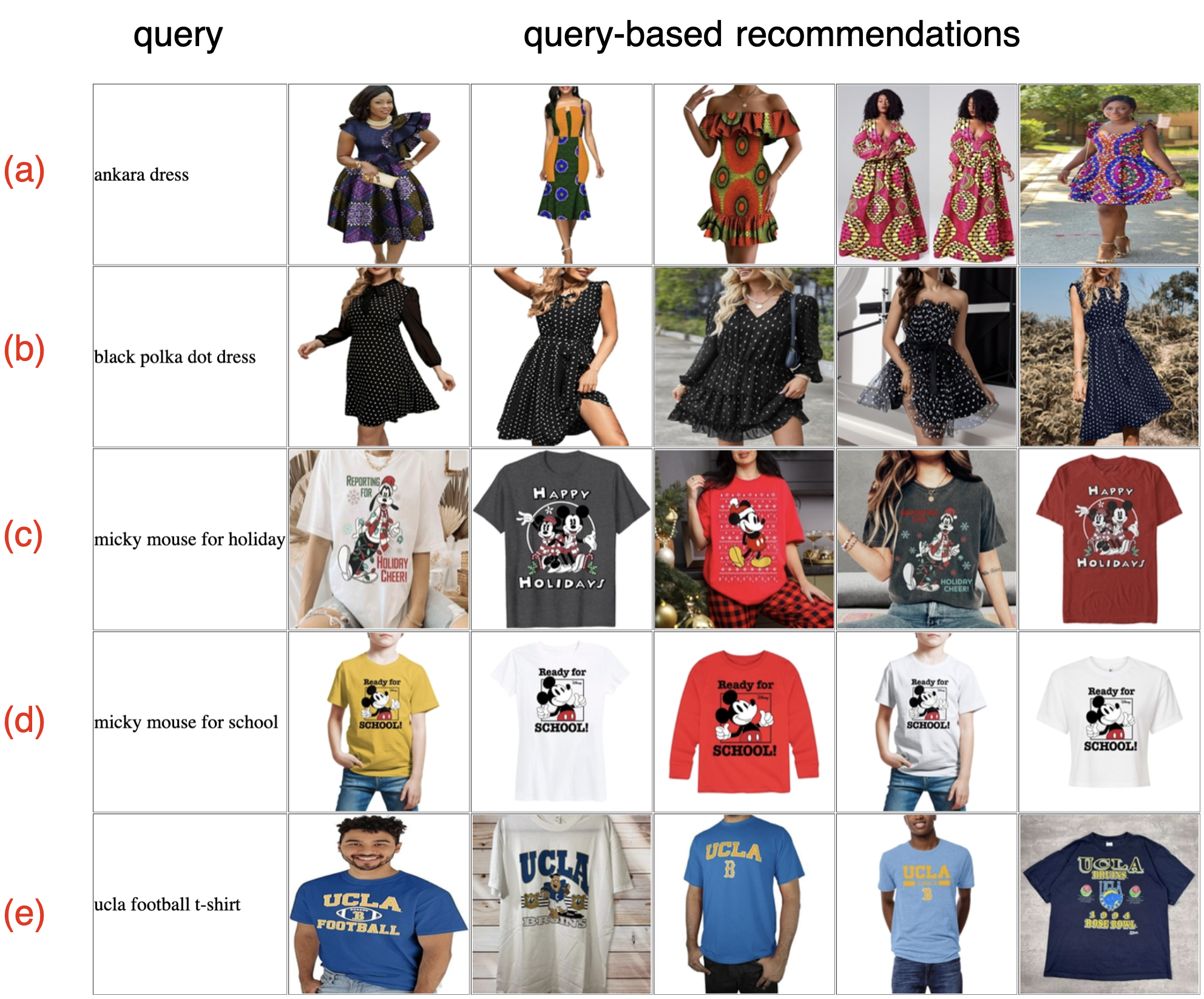}
    \caption{Examples of query-based retrieval for fashion items: the first column is the query, the rest are top 5 recommended items.  (a) Recommendations for the query \textit{``ankara dress''},  (b) Recommendations for the query \textit{``black polka dot dress''},  (c) Recommendations for the query \textit{``mickey mouse for holiday''},   (d) Recommendations for the query \textit{``mickey mouse for school''},  (e) Recommendations for the query \textit{``UCLA football t-shirt''}.}
    \Description{}
    \label{fig:query_fashion}
\end{figure*}

\begin{figure*}[h]
    \centering
    \includegraphics[width=0.98\textwidth]{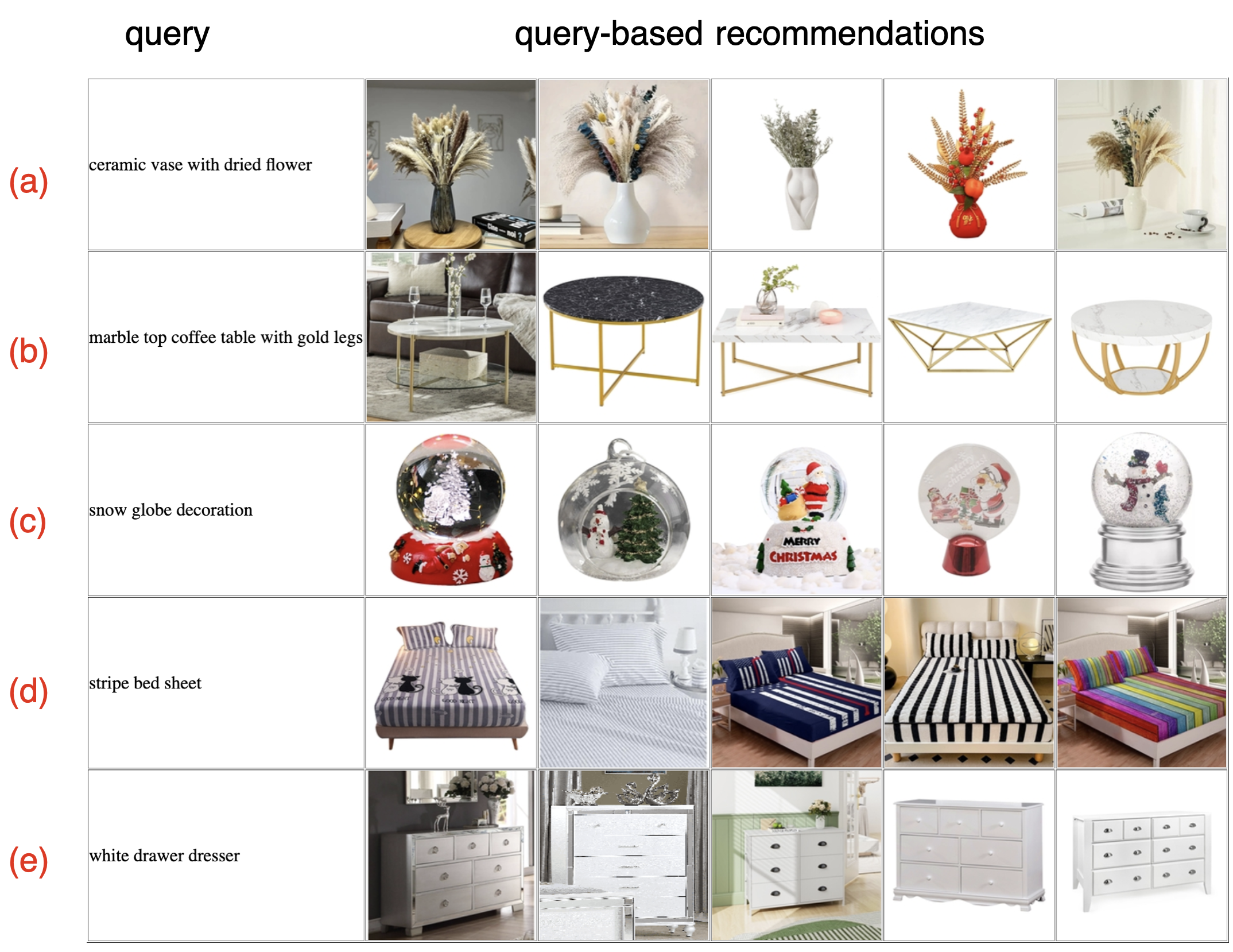}
    \caption{Examples of query-based retrieval for home items: the first column is the query, the rest are top 5 recommended items. 
    (a)  Recommendations for the query \textit{``ceramic vase with dried flower''},  (b)  Recommendations for the query \textit{``marble top coffee table with gold legs''},   (c)  Recommendations for the query \textit{``snow globe decoration''},  (d)  Recommendations for the query \textit{"stripe bed sheet"},  (e)  Recommendations for the query \textit{``white drawer dresser''}.}
    \Description{}
    \label{fig:query_home}
\end{figure*}

\begin{figure*}[h]
    \centering
    \includegraphics[width=0.98\textwidth]{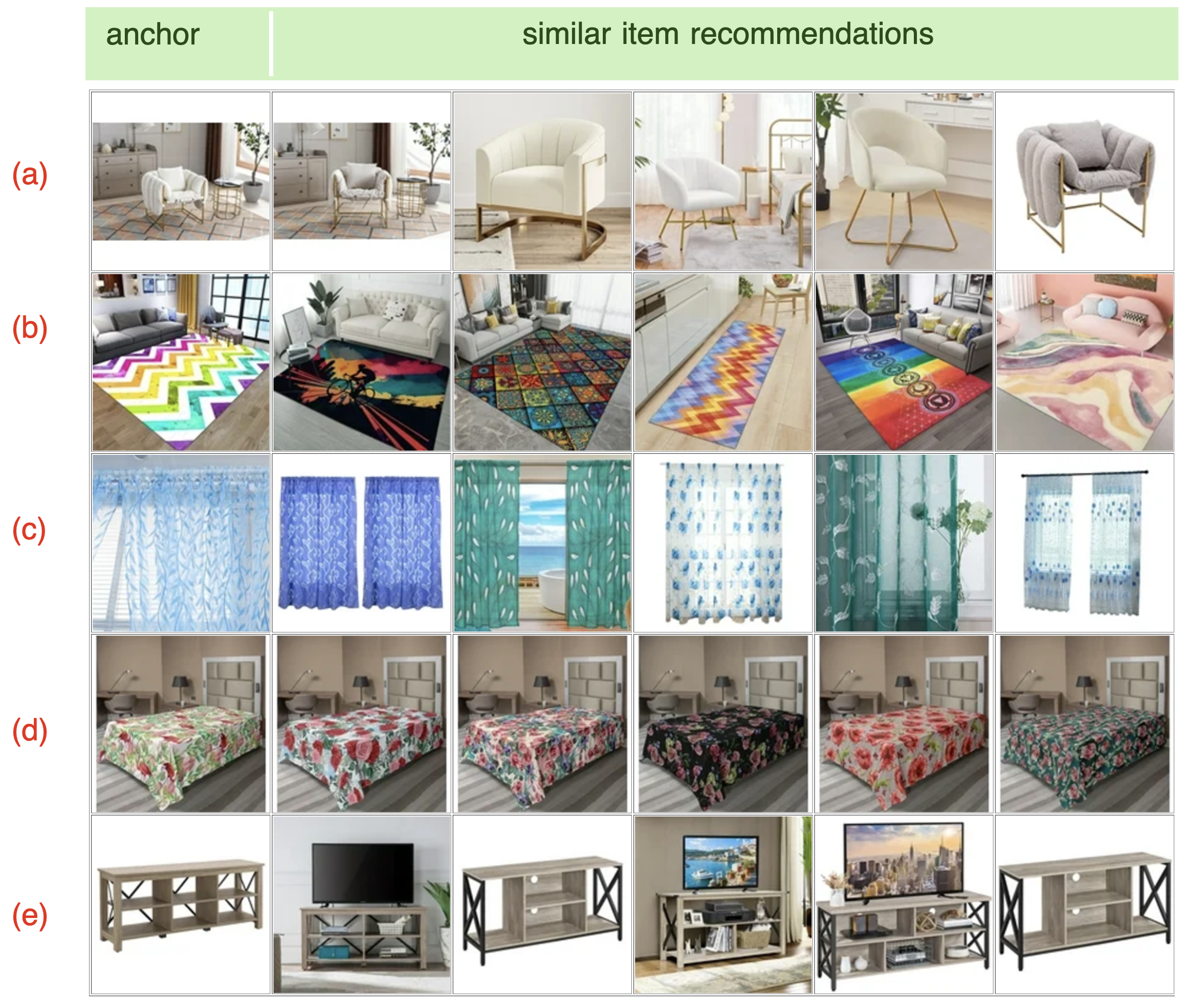}
    \caption{Examples of similar item recommendation for home items: the first column is the anchor item, the rest are top 5 recommendation items based on image similarity. 
    (a) The anchor image is a modern white and gold accent chair. The recommended items share a similar combination of white upholstery with gold or metallic legs, maintaining a contemporary and elegant aesthetic.
 (b) The anchor image is a colorful geometric-patterned area rug. The recommendations feature vibrant color schemes, bold geometric patterns, and similar rug layouts to match the original design.
 (c) The anchor image is a pair of light blue sheer curtains. The recommended items include sheer or semi-sheer curtains with floral, botanical, or abstract patterns, preserving the soft and airy look.
 (d) The anchor image is a floral-patterned bedspread with red and pink roses. The retrieved items emphasize floral patterns with similar color palettes and intricate designs, maintaining a cozy and decorative appearance.
 (e) The anchor image is a wooden TV stand with an open-shelf design and black metal frame. The recommended items feature a similar industrial or rustic style, combining wood surfaces with black metal elements for structural support and aesthetics.}
 \Description{}
    \label{fig:si_home}
\end{figure*}

This section presents visualizations of query-based retrieval and similar item recommendation (SI) tasks for Fashion and Home items. Figures \ref{fig:query_fashion}, \ref{fig:query_home}, and \ref{fig:si_home} illustrate top retrieved results based on text queries and anchor images.

In Figures~\ref{fig:query_fashion} and~\ref{fig:query_home}, each row shows a text query in the first column and the top 5 recommended products in the remaining columns. The fashion queries range from specific clothing types (e.g., “ankara dress,” “UCLA football t-shirt”) to themed queries like “mickey mouse for school.” Home-related queries include decor and furniture items such as “marble top coffee table with gold legs” and “stripe bed sheet.” The results reflect the model’s ability to capture fine-grained semantic details from text.

Figure~\ref{fig:si_home} shows similar item recommendations for home products, where each anchor image is followed by visually similar items. Examples include accent chairs, patterned rugs, bedspreads, and TV stands. The recommended items closely match the anchors in terms of material, color scheme, and overall style, highlighting the model’s effectiveness in image-based similarity retrieval.

Together, these examples demonstrate VL-CLIP’s strength in both multimodal understanding and visual matching across product categories.

\clearpage

\clearpage

\clearpage
\section{Training Algorithm}
\label{appendix:training_algo}
Algorithm~\ref{alg:VL-CLIP} outlines the step-by-step process for building the VL-CLIP model, including the steps for constructing the image/text pairs, localization, query refinement, and finally fine-tuning the model.

\begin{algorithm}[h!]
\caption{VL-CLIP Algorithm} 
\label{alg:VL-CLIP} 
\begin{algorithmic}[1]

\For{each product $n$}
    \State $I_{\text{norm}} \gets \text{ResizeAndNormalize}(I_n)$
    \State $t_{\text{concat}} \gets [t_p \parallel t_g \parallel t_{\text{raw}} \parallel t_{\text{in-context}}]$
    \State Store $I_{\text{norm}},t_{\text{concat}}$
\EndFor
\For{each $I_{\text{norm}}$}

    \State $(B, s) \gets \text{GroundingDINO}(I_{\text{norm}}, t_p)$
    \State \textbf{if} $\max(s) \geq \tau_{\text{thresh}}$ \textbf{then} $I_{\text{crop}} \gets \text{Crop}(I_{\text{norm}}, \arg\max(s))$
    \State \textbf{else} $I_{\text{crop}} \gets I_{\text{norm}}$
    \State Append $I_{\text{crop}}$ to $I_{\text{refined}}$
\EndFor
\For{each $t_{\text{concat}}$}
    \State ${q^{\text{init}}}$$ \gets \text{Summarizer}(t_{\text{concat}})$
    \State $ {q^0} \gets {q^{\text{init}}}$
    \For{$i = 1$ \textbf{to} $5$}
        \State $e^i \gets \text{Evaluator}(q^{i-1}, t_{\text{concat}})$
        \If{\texttt{<STOP>} in $e^i$}
            \State \textbf{break}
        \Else
            \State $q^i \gets \text{Refiner}(q^{i-1}, e^i, t_{\text{concat}})$
            \State $i \gets i + 1$
        \EndIf
    \EndFor
    \State Add $q^{i-1}$ to query set $Q$
\EndFor
\For{$\text{epoch} = 1$ to $N_{\text{epochs}}$}
    \For{each batch of items from $I_{\text{refined}}, Q$}
        \State Compute $(\phi_{\text{CLIP-image}}, \phi_{\text{CLIP-text}})$
        \State Compute $\mathcal{L}_{\text{CLIP}}$ and update parameters via gradient descent
    \EndFor
\EndFor

\end{algorithmic} 
\end{algorithm}

\subsection{Agent Prompts}
\label{sec:synthesis_prompts}
In this section, we list our system prompts in Table \ref{tab:agentSystemPrompts} and user prompts in Table \ref{tab:agentUserPrompts}

\subsection{LLM-Driven Textual Query Synthesis Examples}
\begin{center}
    \includegraphics[width=0.4\linewidth]{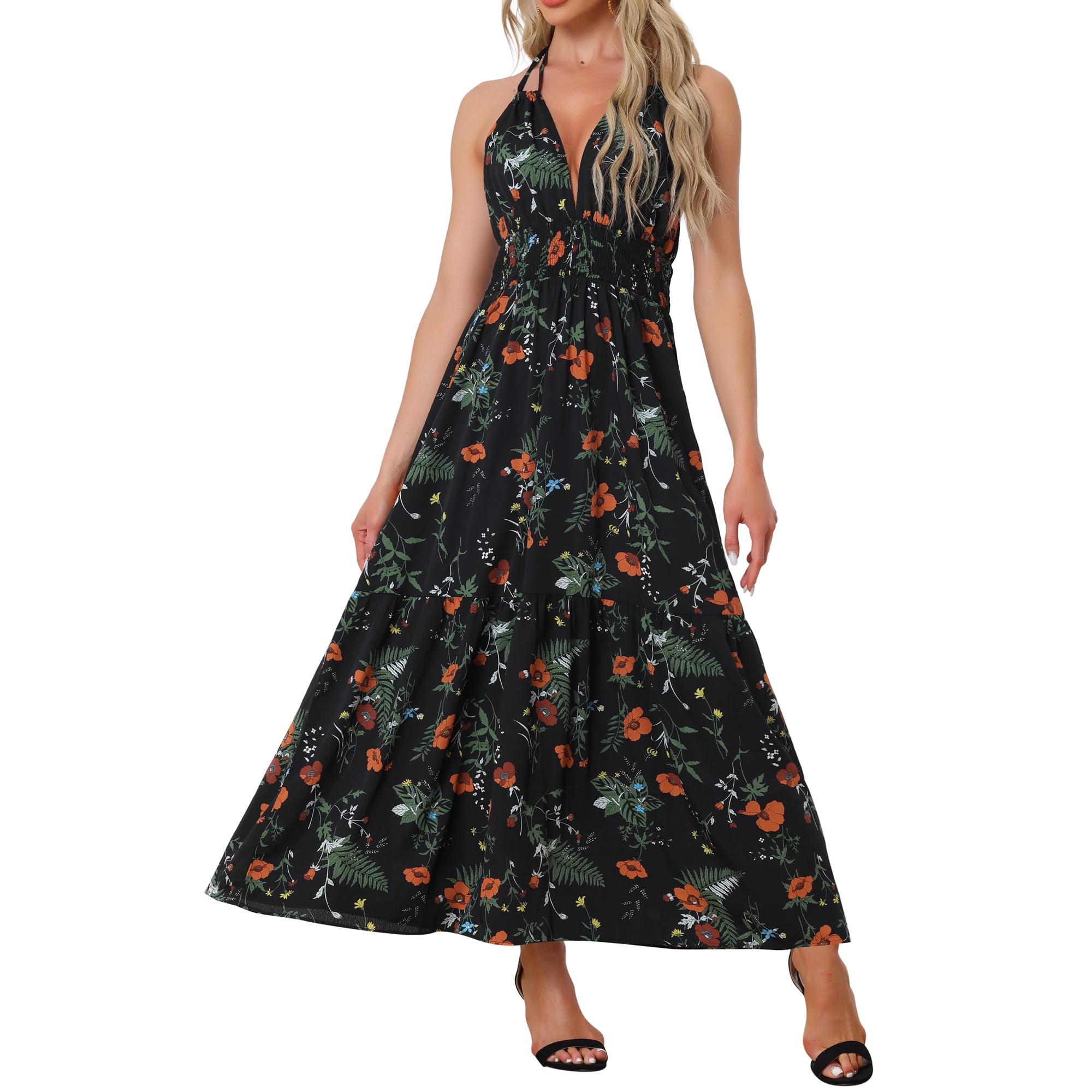}
\end{center}

\begin{Verbatim}[breaklines=true, breakanywhere=true, breaksymbol={}]
"productName": "Unique Bargains Women's Sleeveless Halter Neck Boho Elastic Waist Floral Maxi Sundress S Black",
"productLongDesc": "Machine Wash Cold with Like Colors. Floral. Regular. Long. Sleeveless. Boho. 100% Polyester. Imported",
"productShortDesc": "This floral boho dress is made of several design point: floral pattern, halter,sleeveless, beach maxi dress. Styled with the earings, hand bag and high heels as your chic look! Floral Maxi Dress, Boho Dress, Halter Neck, Elastic Waist. Suit for summer and many occasion,such as beach, party, vacation dating. MODEL SHOW: Regular Fit. Wear in size XS. Wash Cold with Like Color.. ",
"productType": "Dresses"
\end{Verbatim}

\noindent\textbf{\color{dullorange} \texttt{<Initial Summary>}} 

\begin{Verbatim}[breaklines=true, breakanywhere=true, breaksymbol={}]
Sleeveless halter neck floral maxi sundress with elastic waist, black, boho style, 100% polyester.
\end{Verbatim}

\noindent\textbf{\color{dullorange} \texttt{<Evaluator iteration 1>}} 

\begin{Verbatim}[breaklines=true, breakanywhere=true, breaksymbol={}]
Score: 4/5
Justification: The summary is concise and captures key visual elements such as the sleeveless halter neck, floral pattern, maxi length, elastic waist, and boho style. However, the inclusion of the material composition (100% polyester) is not visually observable.

Suggestions:
1. Remove the information of [100% polyester].
\end{Verbatim}

\noindent\textbf{\color{dullorange} \texttt{<Refiner iteration 1>}} 

\begin{Verbatim}[breaklines=true, breakanywhere=true, breaksymbol={}]
Sleeveless halter neck floral maxi sundress with elastic waist, black, boho style.
\end{Verbatim}

\noindent\textbf{\color{dullorange} \texttt{<Evaluator iteration 2>}} 

\begin{Verbatim}[breaklines=true, breakanywhere=true, breaksymbol={}]
Score: 5/5
Justification: The summary is concise and captures key visual elements such as the sleeveless halter neck, floral pattern, maxi length, elastic waist, and boho style. It also includes the color (black), which is visually observable. No non-visual details are included, and the summary is less than 20 words.
Suggestions: 5. Do nothing.
<STOP>
\end{Verbatim}

\begin{center}
    \includegraphics[width=0.5\linewidth]{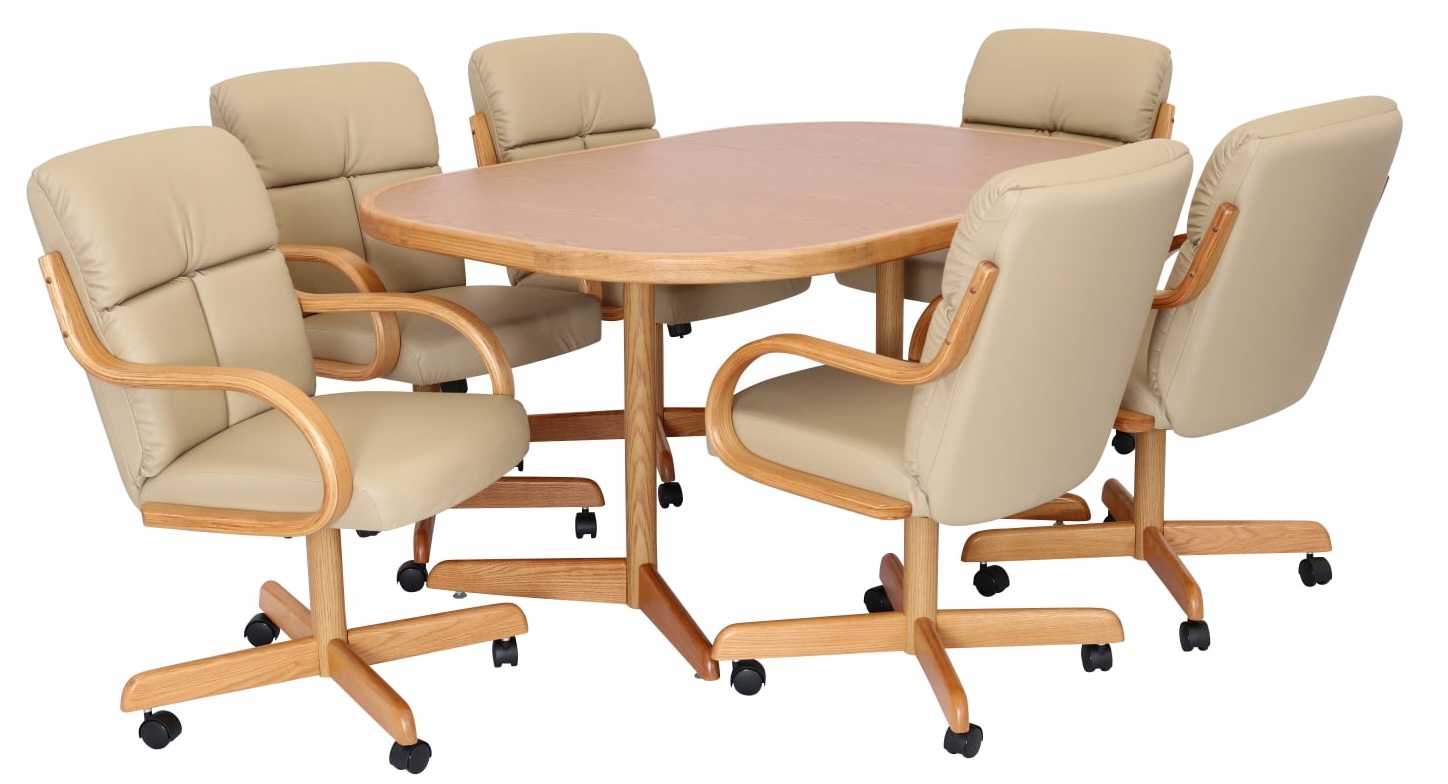}
\end{center}

\begin{Verbatim}[breaklines=true, breakanywhere=true, breaksymbol={}]
"productName": "Caster Chair Company 7-Piece 42x[42/60] Caster Dining Set Wood Edge Laminate Table Top & Desert Chairs",
"productLongDesc": "DURABLE AND STYLISH: This 7-piece dining set features a solid oak wood edge laminate table top with a steel base, paired with 6 caster chairs upholstered in desert polyurethane and oak-finished plywood arms. COMFORTABLE SEATING: The 28\"D x 23\"W x 36\"H caster chairs have a 360-degree swivel and tilt mechanism, making it easy to move and adjust your seating position. Each chair has a weight ....",
"productShortDesc": "Introducing Frankie 7-Piece Caster Dining Set by Caster Chair Company \u2013 a stunning combination of durability, style, and versatility! ",
"productType": "Dining Furniture Sets"
\end{Verbatim}

\noindent\textbf{\color{dullorange} \texttt{<Initial Summary>}} 

\begin{Verbatim}[breaklines=true, breakanywhere=true, breaksymbol={}]
Seven-piece dining set with oak wood edge laminate table, steel base, and six desert polyurethane caster chairs with oak-finished arms.
\end{Verbatim}

\noindent\textbf{\color{dullorange} \texttt{<Evaluator iteration 1>}} 

\begin{Verbatim}[breaklines=true, breakanywhere=true, breaksymbol={}]
Score: 4/5
Justification: The summary is concise and captures key visual elements such as the oak wood edge laminate table, steel base, and desert polyurethane caster chairs with oak-finished arms. However, it could include more specific details about the table's extendable feature and the solid oak wood crowns on the chair and table bases.
Suggestions:
1. Add the information of [extendable bow-end table top].
2. Add the information of [solid oak wood crowns on chair and table bases].
\end{Verbatim}

\noindent\textbf{\color{dullorange} \texttt{<Refiner iteration 1>}} 

\begin{Verbatim}[breaklines=true, breakanywhere=true, breaksymbol={}]
Seven-piece dining set with extendable bow-end oak wood edge laminate table, steel base, and desert polyurethane caster chairs with oak-finished arms and solid oak wood crowns.
\end{Verbatim}

\noindent\textbf{\color{dullorange} \texttt{<Evaluator iteration 2>}} 

\begin{Verbatim}[breaklines=true, breakanywhere=true, breaksymbol={}]
Score: 4/5
Justification: The summary is concise and captures key visual elements such as the extendable bow-end oak wood edge laminate table, steel base, and desert polyurethane caster chairs with oak-finished arms and solid oak wood crowns. However, it is slightly over the word limit.
Suggestions:
1. Shorten the summary.
\end{Verbatim}

\noindent\textbf{\color{dullorange} \texttt{<Refiner iteration 2>}} 

\begin{Verbatim}[breaklines=true, breakanywhere=true, breaksymbol={}]
Seven-piece dining set with oak wood edge laminate table, steel base, and desert polyurethane caster chairs with oak-finished arms.
\end{Verbatim}

\noindent\textbf{\color{dullorange} \texttt{<Evaluator iteration 3>}} 

\begin{Verbatim}[breaklines=true, breakanywhere=true, breaksymbol={}]
Score: 5/5
Justification: The summary is concise and captures key visual elements such as the oak wood edge laminate table, steel base, and desert polyurethane caster chairs with oak-finished arms.No non-visual details are included, and the summary is less than 20 words.
Suggestions: 5. Do nothing.
<STOP>
\end{Verbatim}

\begin{table*}[h!]
    \caption{System Prompts for Summarizer, Evaluator, and Refiner {Agents}}
    \centering
    \begin{tabularx}{\textwidth}{p{3cm} X}
    \hline
    \textbf{{Agent}} & \textbf{System Prompt} \\
    \hline
    \textbf{Summarizer} 
        & You are a product copywriter, skilled in creating concise and visually-rich summaries. Your task is to generate a less than 20-words description that vividly encapsulates the product’s visual observable elements, without using sales language.

        \textbf{Instructions:}
        \begin{itemize}
            \item Limit the description to less than 20 words.
            \item Concentrate on capturing visually observable attributes such as color, texture, shape, and material.
            \item Refrain from using sales or persuasive language.
        \end{itemize}
        \\ 
    \hline
    \textbf{Evaluator}
        & You are a summary evaluator for product copywriting. Your task is to evaluate a product summary according to the following criteria:

        \textbf{Instructions:}
        \begin{itemize}
            \item The summary must be less than 20 words.
            \item It must encapsulate the product’s visually observable elements (such as color, texture, shape, material).
            \item It must refrain from using sales or persuasive language.
            \item It must not include non-visual details such as prices, brand names, benefits, or any abstract descriptors.
            \item Only ask for the information that appeared in the product details.
            \item Provide feedback and revision suggestions focused on the presence or absence of visual elements only.
        \end{itemize}

        You give scores for the summaries, justification for the scores as well as revise suggestions.
        Your score should correspond to your suggestions. Your suggestions can be:
        (1) Add the information
        (2) Remove the information 
        (3) Rephrase the information 
        (4) Shorten the summary.
        (5) Do nothing.

        If you find the summary is too long, ask for a short summary.
        If the summary includes any non-visual content, instruct to remove it. Only consider information that is present in the product details and is visually observable. If you determine that no further revisions are needed, end your output with \texttt{"<STOP>"} (without any extra text). 
        \\
    \hline
    \textbf{Refiner}
        & You are a skilled product copywriter, experienced in creating concise and visually-rich summaries.
        Your task is to refine the summary. Follow all the suggestions and you can not make more comments. 
        Give one final summary as output.

        \textbf{Instructions:}
        \begin{itemize}
            \item Use only details present in the product data.
            \item Exclude any information not found in the product details.
            \item Limit the summary to fewer than 20 words.
            \item Focus solely on visually observable attributes: color, texture, shape, and material.
            \item Do not include measurements, prices, brand names, or benefits.
            \item Provide one final refined summary with no additional commentary.
            \item Do not include any extra text or a revised summary in your output.
        \end{itemize}
        \\
    \hline
    \end{tabularx}
    \label{tab:agentSystemPrompts}
\end{table*}

\begin{table*}[h!]
    \caption{User Prompts for Summarizer, Evaluator, and Refiner {Agents}}
    \centering
    \begin{tabularx}{\textwidth}{p{3cm} X}
    \hline
    \textbf{{Agent}} & \textbf{User Prompt} \\
    \hline
    \textbf{Summarizer} & 
Product Details: \{Product Details\} 
[In-context Examples] 
    \\
    \hline
    \textbf{Evaluator} &
    Please evaluate the product summary below in light of the product details provided.\newline
    Product Details: \{Product Details\} \newline
    Summary Content: \{Summary Content\} \newline
    The output should be a probability distribution of assigning the score between 1-5 as well as its justification.\newline
    Please provide comments if you think this summary is not good enough. \newline
    [In-context Examples] \{Memory\}
    \\
    \hline
    \textbf{Refiner} &
    Please refine the summary based on the following details:\newline
    Product Details: \{Product Details\} \newline
    Summary Content: \{Summary Content\} \newline
    Evaluator Feedback: \{Evaluator Feedback\} \newline
    Provide one final summary as output. \newline
    [In-context Examples] \{Memory\} \\
    \hline
    \end{tabularx}
    \label{tab:agentUserPrompts}
\end{table*}

\section{Deployment Algorithm}
\label{appendix:deployment_algo}

This section describes the deployment algorithm for the VL-CLIP framework, providing the steps for scalable processing, embedding generation, and efficient retrieval using the HNSW index. 

\begin{algorithm}[h!]
\caption{VL-CLIP Framework: Deployment}
\label{alg:VL-CLIPd}
\begin{algorithmic}[1]
\State $\mathcal{D}_{\mathrm{hash}} \gets \varnothing,\quad \mathcal{I}_{\mathrm{unique}} \gets \varnothing$
\For{each $I \in \mathcal{I}_{\text{refined}}$}
   \State $h_{\mathrm{phash}} \gets \text{PerceptualHash}(I)$
   \If{$h_{\mathrm{phash}} \notin \mathcal{D}_{\mathrm{hash}}$}
       \State $\mathcal{D}_{\mathrm{hash}} \gets \mathcal{D}_{\mathrm{hash}} \cup \{h_{\mathrm{phash}}\}$
       \State $\mathcal{I}_{\mathrm{unique}} \gets \mathcal{I}_{\mathrm{unique}} \cup \{I\}$
  \EndIf
\EndFor

\State Partition $\mathcal{I}_{\mathrm{unique}}$ into batches $\{\mathcal{B}_1, \,\mathcal{B}_2, \dots\}$
\For{each $\mathcal{B}_i$ \textbf{in parallel}}
   \For{each $I \in \mathcal{B}_i$}
       \State $\mathbf{v}_I \gets \phi_{\text{CLIP-image}}(I)$
  \EndFor
   \State Store $\{\mathbf{v}_I\}$ in embedding repository
\EndFor

\vspace{0.5em}
\State $\mathcal{H} \gets \text{BuildHNSW}\bigl(\{\mathbf{v}_I \,\mid\, I\in\mathcal{I}_{\mathrm{unique}}\}\bigr)$

\State Given query $q$:
\State \quad $\mathbf{t}_q \gets \phi_{\text{CLIP-text}}(q)$
\State \quad $R_{\mathrm{ANN}} \gets \mathcal{H}\,\text{.search}(\mathbf{t}_q,\,K)$
\State \quad \textbf{return} top-$K$ items in $R_{\mathrm{ANN}}$

\end{algorithmic}
\end{algorithm}

\section{{VLM} Evaluation Process}
\label{appendix:eval_process}

\begin{figure*}[h]
    \centering
    \includegraphics[width=0.90\linewidth]{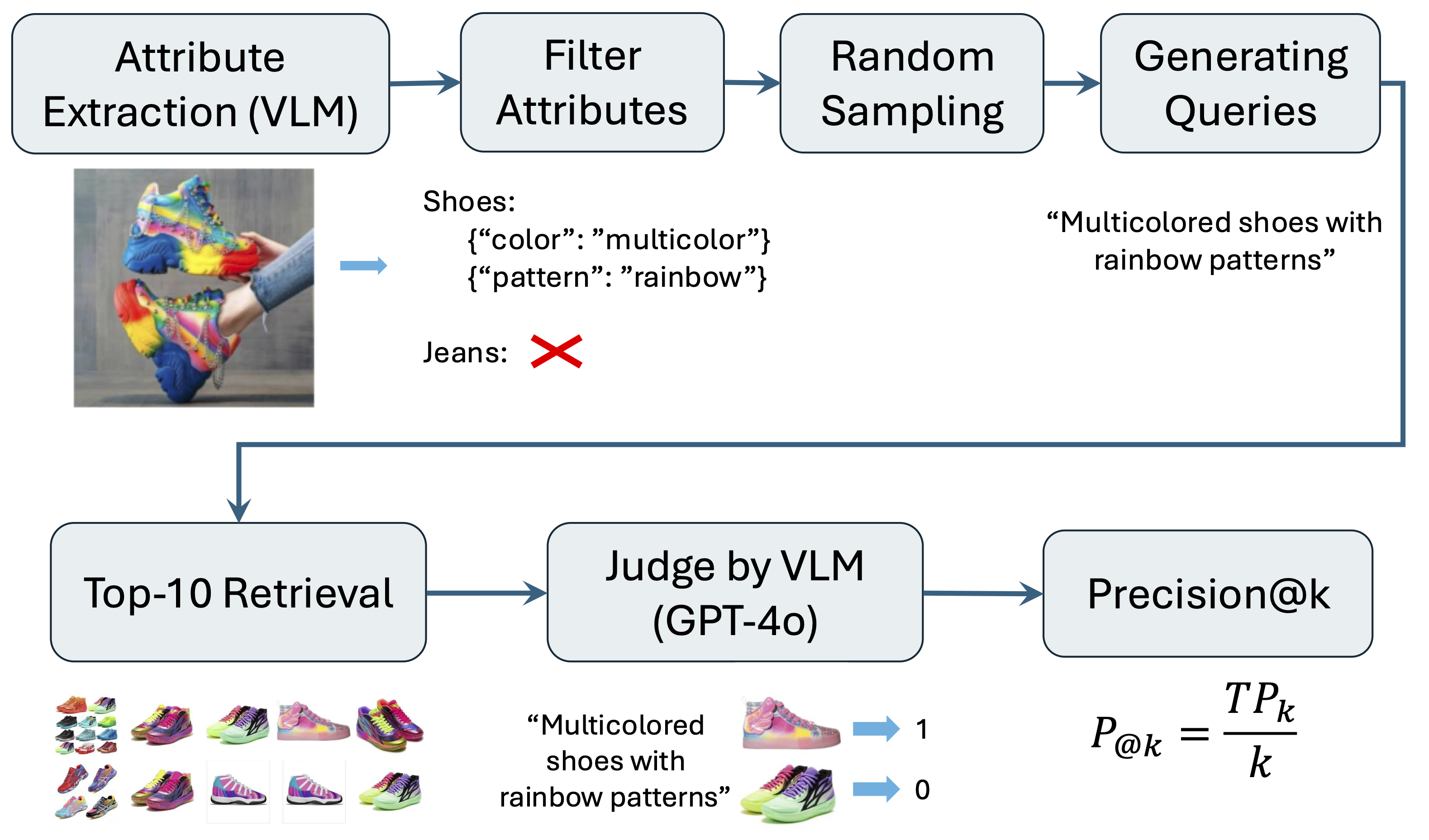}
    \caption{Query-based evaluation process using {VLM}}
    \Description{}
    \label{fig:llm_eval_diagram}
\end{figure*}

We assess the effectiveness of our query-based retrieval approach on \textit{Walmart.com E-Commerce Dataset} using an { VLM}-based evaluation framework. Our methodology follows a structured pipeline and consists of \textbf{automated query generation}, and \textbf{{VLM}-as-judge evaluation}, as described in Figure \ref{fig:llm_eval_diagram}.

\subsection{Automated Query Generation}
\begin{itemize}
\item \textbf{Attribute extraction:} We apply a Vision-Language Model (VLM) to extract structured attributes from a random subset of  product items.  Given an input image, the extracted attributes an be represented as 
$$A=\{(a_1, v_1), (a_2, v_2), ..., (a_m, v_m)\}$$
where $a_i$ represents an attribute type (e.g., ``\textit{color}'') and $v_i$ is its value (e.g., ``\textit{blue}'', or ``\textit{multicolor}'').
The extracted attributes are filtered to ensure they are directly relevant to the primary item in the image, resulting in $A_{filtered}$.

\item \textbf{Query generation:} We utilize an LLM to generate search queries from extracted attributes. Given the filtered attribute set $A_{filtered}$ for  item $X$, the query is generated by $Q=LLM(A_{filtered})$. For instance, an item from the "T-shirt" products with attributes 
\begin{itemize}
\item \textit{``sleeve\_length'' = ``long''}
\item \textit{``pattern'' = ``Mickey Mouse''}
\item \textit{``pattern\_placement'' = ``front, center''}
\end{itemize}
is transformed into the query: ``\textit{T-shirt with long sleeves and Mickey Mouse pattern on front}''.
\newline This structured approach enables a fair comparison across datasets while ensuring that the generated queries align with real-world search behaviors.
\end{itemize}

\subsection{{VLM}-as-judge Evaluation}
\label{appendix:llm_eval}

\begin{itemize}
\item \textbf{Top-K retrieval:} For each query, we retrieve the top-K results, $R_K$:
$$R_k=\left\{I_1, I_2, ..., I_K\right\}$$
where K=10. The retrieved items are ranked based on their relevance to the query.
\item \textbf{Relevance assessment:} Each retrieved image $I_j$ is paired with its corresponding query and the level of relevance of pair is measured by a VLM (GPT-4o), assigning a binary relevance score:

\[
S(Q, I_j) = 
\begin{cases}
1, & \text{if } I_j \text{ matches query } Q \\
0, & \text{otherwise}
\end{cases}
\]

{ The prompt used for this evaluation is listed in Table~\ref{tab:vlm_prompts}}.

\item \textbf{Performance metrics:} We compute $\text{Precision}@k$ for $k\in\{1,3,5\}$. 
$$\text{Precision}@k = \frac{TP_k}{k}$$
where, $TP_k$ is the number of correctly retrieved relevant items within the top-$k$, and $k$ is the total number of retrieved results. 
\end{itemize}

\subsection{Similar Item Recommendation}
We also evaluate the model's performance through a similar item recommendation task, as follows:
\begin{itemize}
    \item We randomly select $N$ anchor items, where $N=100$. For each anchor, we retrieve the top-K recommendations, where $K\in \{1, 3, 5\}$. 
    \item Each anchor is paired with its recommended items, and we use a large language model (GPT-4o) to assess similarity. The model assigns a binary relevance score (0 or 1) to each anchor-recommendation pair, where 1 indicates a pair is similar and 0 indicates that they are not similar.
    {The specific prompt employed for assessing visual similarity is provided in Table~\ref{tab:vlm_prompts}.}
    \item We use the same performance metrics as in the query-based retrieval approach.
\end{itemize}

Table~\ref{tab:vlm_prompts} shows the prompts used for VLM-as-Judge evaluation.
\begin{table}[H]
    \caption{Prompts used for VLM-as-Judge evaluation}
    \begin{tabular}{p{1.5cm} p{6.0cm}}
        \toprule
        \textbf{{Prompt Type}} & \textbf{Prompt} \\
        \midrule
        \textbf{{Query-Based Retrieval}} &
        Analyze the image and the query below. Answer strictly with 0 or 1 to identify whether the visual characteristics in the image match with the query:
        \begin{itemize}
          \item Return 1 if the visual characteristics of the image match the attributes, product type, and details described in the query,
          \item Return 0 if they do not match.
        \end{itemize}
        \texttt{\{query\} \{image\}}
        \\
        \midrule
        \textbf{{Similar Item Recommendation}} &
        Identify with 0 or 1 whether the two images are similar in terms of their visual characteristics such as pattern, style, design. This is a verification step for a visually similar item recommendation task.

        Example: For two input images of t-shirts that are both round-neck, return 1. But if one image is round-neck and the other is v-neck, return 0.

        As long as the two items are from the same product type and some of the main characteristics (pattern, style, design) of the two products are similar, provide 1; otherwise provide 0.

        \texttt{\{image1\} \{image2\}}
        \\
        \bottomrule
    \end{tabular}
    \label{tab:vlm_prompts}
\end{table}

{
\section{Cross-Domain Generalization}
\label{appendix:generalizability}

To assess the generalizability of our approach, we extend our experiments beyond the original domains by evaluating additional categories including Walmart Art and Toys under zero-shot settings.

Table~\ref{tab:multi-modal-retrieval-performance-art-and-toy} reports zero-shot multi-modal retrieval results on Art and Toy categories. We observe that VL-CLIP consistently outperforms other models, demonstrating strong transferability to new product types.
}
    
\begin{table}[htbp]
    \centering
    \caption{Zero-shot multi-modal retrieval performance of different models on Art and Toy datasets} 
    \begin{tabular}{l cc cc}
        \toprule
        \multirow{2}{*}{\textbf{Model}} & \multicolumn{2}{c}{\textbf{Art}} & \multicolumn{2}{c}{\textbf{Toy}} \\
        \cmidrule(lr){2-3} \cmidrule(lr){4-5}
        & \textbf{HITS@5} & \textbf{MRR} & \textbf{HITS@5} & \textbf{MRR} \\
        \midrule
    
        CLIP         & 0.3287 & 0.2319 & 0.3442 & 0.2625 \\
        FashionCLIP  & 0.1405 & 0.0972 & 0.3981 & 0.2283 \\
        GCL          & 0.1660 & 0.1233 & 0.2153 & 0.1586 \\
        VL-CLIP      & \textbf{0.4492} & \textbf{0.3974} & \textbf{0.5175} & \textbf{0.3791} \\
        \bottomrule
    \end{tabular}
    \label{tab:multi-modal-retrieval-performance-art-and-toy}
\end{table}

Table~\ref{tab:llm_evaluation_art_and_toy} shows LLM-based evaluation for both query-based retrieval and similar item recommendation on Walmart's Art and Toy categories.
\begin{table}[H]
    \caption{{VLM}-evaluation results on Walmart Art and Toy categories in zero-shot setting.}
    \renewcommand{\arraystretch}{0.9}
    \begin{tabular}{lccc}
        \toprule
        \multicolumn{4}{c}{\textbf{Query-based retrieval}} \\
        \midrule
        \textbf{Model} & \textbf{Precision@1} & \textbf{Precision@3} & \textbf{Precision@5} \\
        \midrule
        CLIP         & 0.4164 & 0.4458 & 0.4479 \\
        FashionCLIP  & 0.3762 & 0.4235 & 0.4319 \\
        GCL          & 0.3404 & 0.3770 & 0.3906 \\
        VL-CLIP      & \textbf{0.6317} & \textbf{0.6177} & \textbf{0.6267} \\
        \midrule
        \multicolumn{4}{c}{\textbf{Similar item recommendation}} \\
        \midrule
        \textbf{Model} & \textbf{Precision@1} & \textbf{Precision@3} & \textbf{Precision@5} \\
        \midrule
        CLIP         & 0.8565 & 0.8299 & 0.8577 \\
        FashionCLIP  & 0.8477 & 0.7983 & 0.8054 \\
        GCL          & 0.7849 & 0.7476 & 0.7695 \\
        VL-CLIP      & \textbf{0.9340} & \textbf{0.8854} & \textbf{0.8871} \\
        \bottomrule
    \end{tabular}
    \label{tab:llm_evaluation_art_and_toy}
\end{table}

\end{document}